\PassOptionsToPackage{table,xcdraw}{xcolor}
\documentclass[sigconf]{acmart}

\usepackage[shortlabels]{enumitem}
\usepackage{booktabs}
\usepackage{framed,enumitem} 
\usepackage{scalerel}
\usepackage{algorithm}
\usepackage{algpseudocode}
\usepackage[most]{tcolorbox}
\newtcolorbox{nlpseudo}[1][]{colback=gray!7,colframe=gray!45,boxrule=0.3pt,arc=2pt,left=6pt,right=6pt,top=6pt,bottom=6pt,#1}
\usepackage[utf8]{inputenc} 
\usepackage{newunicodechar}
\usepackage{float} 
\definecolor{LightBlueGrey}{RGB}{220,230,240} 

\renewcommand{\arraystretch}{1.15}

\usepackage{array,ragged2e,tabularx,booktabs,graphicx}

\renewcommand{\arraystretch}{1.15}
\newcolumntype{Y}{>{\RaggedRight\arraybackslash\hspace{0pt}\vspace{0pt}}X}

\newcommand{\codewrap}[1]{\ttfamily\RaggedRight\hspace{0pt}#1}

\usepackage{placeins} 
\usepackage{float}

\newunicodechar{→}{\ensuremath{\rightarrow}}
\newunicodechar{↔}{\ensuremath{\leftrightarrow}}
\newunicodechar{∈}{\ensuremath{\in}}
\newunicodechar{≥}{\ensuremath{\ge}}
\newunicodechar{≤}{\ensuremath{\le}}
\newunicodechar{×}{\ensuremath{\times}}
\newunicodechar{−}{\ensuremath{-}} 
\newunicodechar{…}{\ldots}
\usepackage[most]{tcolorbox}
\usepackage{xstring}   
\usepackage{pgf}       

\definecolor{ScoreBlue}{HTML}{1170AA}
\definecolor{ScoreYellow}{HTML}{F2B134}
\definecolor{ScoreRed}{HTML}{E54D37}
\colorlet{ScoreBlueBG}{ScoreBlue!10}
\colorlet{ScoreYellowBG}{ScoreYellow!10}
\colorlet{ScoreRedBG}{ScoreRed!10}

\newcommand{\makeScoreBadge}[3]{%
  \raisebox{-0.2ex}{
    \tcbox[
      on line,
      colback=#2, colframe=#2,
      boxrule=0pt, arc=2pt,
      left=2pt, right=2pt, top=0.2ex, bottom=0.2ex,
      boxsep=0pt,
      nobeforeafter
    ]{\textcolor{#1}{\textbf{#3}}}%
  }%
}

\newcommand{\scorebadge}[1]{%
  \begingroup
    \def\raw{#1}%
    \IfEndWith{\raw}{\%}{\StrGobbleRight{\raw}{1}[\rawnum]}{\def\rawnum{\raw}}%
    \pgfmathparse{\rawnum}\let\num\pgfmathresult
    \def\fg{ScoreBlue}\def\bg{ScoreBlueBG}%
    \pgfmathparse{\num<33?1:0}\ifnum\pgfmathresult=1
      \def\fg{ScoreRed}\def\bg{ScoreRedBG}%
    \else
      \pgfmathparse{\num<=65?1:0}\ifnum\pgfmathresult=1
        \def\fg{ScoreYellow}\def\bg{ScoreYellowBG}%
      \fi
    \fi
    \makeScoreBadge{\fg}{\bg}{#1}%
  \endgroup
}
\usepackage{fontawesome5}
\usepackage{pgf}      

\definecolor{StarGold}{HTML}{F2B134}
\colorlet{StarEmpty}{black!25}

\newlength{\starsep}         
\newlength{\starendtrim}     
\newcommand{\tightstar}[2]{
  \raisebox{-0.1ex}{\rlap{\textcolor{#1}{#2}}}\kern\starsep
}

\newcommand{\starfull}{\tightstar{StarGold}{\faStar}}
\newcommand{\starhalf}{\tightstar{StarGold}{\faStarHalfAlt}}
\newcommand{\starempty}{\tightstar{StarEmpty}{\faRegStar}}

\newcommand{\starrating}[2][]{%
  \begingroup
    \def\starsize{\normalsize}%
    \ifx\relax#1\relax\else\def\starsize{#1}\fi
    \pgfmathtruncatemacro{\full}{floor(#2)}%
    \pgfmathtruncatemacro{\half}{(#2-\full)>=0.5?1:0}%
    \pgfmathtruncatemacro{\empty}{5-\full-\half}%
    {\starsize
      \count0=\full \loop\ifnum\count0>0 \starfull\advance\count0 by -1\repeat
      \ifnum\half>0 \starhalf\fi
      \count0=\empty \loop\ifnum\count0>0 \starempty\advance\count0 by -1\repeat
      \kern-\starsep\kern-\starendtrim\relax
    }%
  \endgroup
}

\newenvironment{tightitemize}{\begin{itemize}\setlength\itemsep{-1.2pt}}{\end{itemize}}

\usepackage{xspace}
\newcommand{\lexara}{\textsc{Lexara}\xspace}
\newcommand{\pheading}[1]{\vspace{4px}\noindent\textbf{#1}}

\usepackage{subcaption}

\AtBeginDocument{%
  \providecommand\BibTeX{{%
    \normalfont B\kern-0.5em{\scshape i\kern-0.25em b}\kern-0.8em\TeX}}}

\copyrightyear{2026}
\acmYear{2026}
\setcopyright{cc}
\setcctype{by}
\acmConference[CHI '26]{Proceedings of the 2026 CHI Conference on Human Factors in Computing Systems}{April 13--17, 2026}{Barcelona, Spain}
\acmBooktitle{Proceedings of the 2026 CHI Conference on Human Factors in Computing Systems (CHI '26), April 13--17, 2026, Barcelona, Spain}
\acmPrice{}
\acmDOI{10.1145/3772318.3790735}
\acmISBN{979-8-4007-2278-3/2026/04}

\begin{document}
\newcolumntype{Y}{>{\RaggedRight\arraybackslash}X}

\title{\lexara: A User-Centered Toolkit for Evaluating Large Language Models for Conversational Visual Analytics}

 \author{Srishti Palani}
 \email{srishti.palani@salesforce.com}
 \affiliation{%
  \institution{Tableau Research, Salesforce}
    \city{Palo Alto, CA}
  \country{USA}
}

\author{Vidya Setlur}
\email{vsetlur@salesforce.com}
\affiliation{%
  \institution{Tableau Research, Salesforce}
    \city{Palo Alto, CA}
  \country{USA}
}

\renewcommand{\shortauthors}{Palani and Setlur}

\begin{abstract}
Large Language Models (LLMs) are transforming Conversational Visual Analytics (CVA) by enabling data analysis through natural language. However, evaluating LLMs for CVA remains a challenge: requiring programming expertise, overlooking real-world complexity, and lacking interpretable metrics for multi-format (visualizations and text) outputs. Through interviews with 22 CVA developers and 16 end-users, we identified use cases, evaluation criteria and workflows. We present \lexara, a user-centered evaluation toolkit for CVA that operationalizes these insights into: (i) test cases spanning real-world scenarios; (ii) interpretable metrics covering visualization quality (data fidelity, semantic alignment, functional correctness, design clarity) and language quality (factual grounding, analytical reasoning, conversational coherence) using rule-based and LLM-as-a-Judge methods; and (iii) an interactive toolkit enabling experimental setup and multi-format and multi-level exploration of results without programming expertise. We conducted a two-week diary study with six CVA developers, drawn from our initial cohort of 22. Their feedback demonstrated \lexara's effectiveness for guiding appropriate model and prompt selection.

\end{abstract}

\ccsdesc[500]{Human-centered computing~Visualization systems and tools}
\ccsdesc[500]{Human-centered computing~Interactive systems and tools}
\keywords{Benchmarking, Analytical Conversation, Visual Analytics, Large Language Model Evaluation}

\maketitle
\section{Introduction}
\label{sec:intro}
Recent advances in Large Language Models (LLMs) have enabled a shift toward more natural conversational interactions with data~\cite{openai2024gpt4technicalreport,microsoftcopilot2025, setlur2016eviza}. Increasingly, LLMs are being integrated into \emph{Conversational Visual Analytics (CVA)} tools, allowing users to generate and refine visualizations through natural language~\cite{tableauAgent2025,googleLookerConversationalAnalytics2024,microsoftPowerBIQA2025, maddigan2023chat2vis,dibia2023lida}. This democratizes \emph{Visual Analytics (VA)}, traditionally defined as analytical reasoning facilitated by interactive visual interfaces~\cite{thomascook:2006,card:1999}, by making it accessible to users without programming or analytical expertise. As CVA tools proliferate, it has become imperative for CVA tool developers and end-user analysts to continually evaluate and adapt to a rapidly growing ecosystem of LLMs and system prompts. These choices directly affect system behavior, output quality, and end-user trust~\cite{Beurer-Kellner:2023}. To understand current evaluation practices and identify gaps in existing approaches, we conducted formative studies with practitioners to investigate the following research questions:
\begin{itemize}[label={}]
    \item \textbf{RQ1:} What does \textbf{practitioners' real-world use of CVA} look like?
    \item \textbf{RQ2:} What \textbf{evaluation criteria} do practitioners apply when assessing CVA system outputs?
    \item \textbf{RQ3:} What \textbf{evaluation workflows} do practitioners use for CVA interactions, what challenges do they face, and how well do existing tools address these challenges?
\end{itemize}

Through semi-structured interviews with 22 CVA tool developers and an observational study with 16 end-users (where a browser extension logged real-world CVA interactions), we uncovered significant gaps between practitioner needs and existing approaches. Thematic analysis revealed that real-world CVA usage is inherently multi-turn and multi-format: users engage in iterative conversations where context from earlier exchanges informs later responses, and expect systems to produce integrated text, visualization, and code outputs. Practitioners evaluate both \textit{visualization quality} (e.g., data fidelity, field similarity, chart type, axes, filters and sorting, visual encodings, and interactivity) and \textit{analytical natural language response quality} (e.g., factual grounding, analytical thinking, conversational coherence, and follow-up relevance across turns), emphasizing the need for flexible, multi-granular evaluation that accommodates graded correctness and multiple valid answers. A response may be technically correct but still suboptimal or misleading. For example, a visualization response with swapped axes, choice of \texttt{pie} vs. \texttt{bar} charts, or semantically inferred fields (e.g., \texttt{Profit} vs. \texttt{Revenue-Cost}) may be valid, but still differ from expected outputs in ways that could affect interpretability and trust in the analyses. Consequently, practitioners rely on ad-hoc, fragmented CVA evaluation workflows: manually comparing outputs across spreadsheets, adapting ill-suited Natural Language Processing (NLP) metrics, and referencing external benchmark reports.

Yet current evaluation approaches fall short of these requirements. Existing CVA benchmarks' test cases~\cite{luo2021nvbench, luo2025nvbench2, chen2024viseval} are synthetically generated, focus primarily on single-turn interactions, and require programming expertise for setup and interpretation, limiting accessibility for product managers, designers, and other low-code stakeholders. Traditional NLP metrics like BLEU~\cite{papineni2002bleu}, ROUGE~\cite{lin2004rouge}, or F1, Precision, Recall~\cite{10.1145/3606367} are limited to n-gram overlap with single references and struggle with multi-format CVA outputs. Even recent visualization-specific metrics~\cite{VISShepherd2025, podo2024vievallmconceptualstack, dibia2023lida, liu2025simvecvisdatasetenhancingmllms} focus on isolated aspects rather than the entire CVA pipeline, and rarely accommodate graded correctness and are difficult to interpret~\cite{zhang-etal-2004-interpreting}. Recent general-purpose LLM evaluation tools~\cite{Kahng2024LLMComparatorTVCG, Arawjo2024ChainForge, Kim2024EvalLM} offer low-code interfaces but lack native support for CVA-specific aspects like rendered visualizations, visualization grammars, analytical thinking in natural language responses, and datasource analysis. These gaps leave practitioners unable to systematically evaluate the LLM-mediated interaction components (e.g., conversational coherence, inferred assumptions, field selection, and visualization correctness) that critically shape whether downstream analytical reasoning is even possible. While visual analytics (VA) research traditionally focuses on assessing end-to-end sensemaking effectiveness, cognitive support, and task-level analytical outcomes~\cite{plaisant2004challenge, kang2009evaluating, pirolli2005sensemaking}, evaluating these LLM-mediated layers requires different approaches that complement rather than replace traditional user-centered VA evaluation.

To address these gaps, we present \lexara, a user-centered toolkit for evaluating LLMs for CVA that operationalizes our formative findings into: 
\begin{itemize}
\item \textbf{CVA test cases} derived from logged real-world end-user analyst interactions (RQ1), 
\item \textbf{Interpretable, graded CVA evaluation metrics} aligned with practitioners' evaluation criteria (RQ2), and
\item \textbf{An interactive low-code CVA-specific benchmarking tool} designed around practitioners' workflow challenges and needs (RQ3), enabling multi-format (text-to-spec visualizations, text-based analytical explanations), multi-turn evaluation and systematic comparison of model-prompt configurations.
\end{itemize}

We deployed \lexara and engaged six of the 22 CVA tool developers in a two-week diary study. Feedback demonstrated that \lexara's test cases capture real-world complexity, offer more interpretable metrics than traditional approaches, and enable practitioners to uncover performance patterns, diagnose model and prompt behavior, and make informed deployment decisions. The toolkit is publicly available at \url{https://lexara-6b38293fcdac.herokuapp.com/} with open-source code at \url{https://anonymous.4open.science/r/Lexara-CVA-Eval-280B/README.md}. Overall, this work makes progress toward the larger vision of responsible AI development for analytics~\cite{holstein2019improving,yang2020re}, enabling practitioners to systematically evaluate, compare, and improve LLM-based systems before deployment.

\section{Related Work}
\label{sec:related}
This paper builds on prior research across three themes: (1) CVA tools, which examine how users generate visualizations via natural language dialogue; (2) CVA evaluation tools, which offer frameworks and interfaces to systematically assess conversational outputs for VA; and (3) visualization and analytical language evaluation methods, which propose both quantitative and qualitative metrics to judge the quality of generated visualizations and analytical explanations. 

\subsection{CVA Tools}
A growing body of work has explored \textit{Conversational Visual Analytics (CVA) tools, i.e., systems that enable users to interact with data and create visualizations through natural language dialogue}~\cite{shen2022towards,voigt2022and}. These tools are designed to lower the technical barriers to data exploration by allowing users to issue queries in natural language, which the tool interprets to: retrieve relevant data fields, select appropriate chart types, assign encodings, and generate visualizations. 

Early CVA tools used keyword recognition and clarifications~\cite{setlur2016eviza, yu2019flowsense, srinivasan2021snowy}, interactive widgets~\cite{sun2010articulate,gao2015datatone}, and gesture-based input~\cite{srinivasan2017orko} during the conversation so that users can intuitively interact with their data without technical or programming expertise. While these interaction mechanisms help, intent inference, deeper analysis, and conversational coherence across turns, ambiguity resolution, etc. remained open challenges~\cite{tory2019mean}. 

With recent advances in LLMs, there has been a marked shift toward more expressive and capable CVA tools that can comprehend colloquial, flexible queries, and generate diverse output formats, including structured code, visualization specifications, rendered charts, and natural language explanations. Tools like Chat2VIS~\cite{maddigan2023chat2vis} leverage GPT-3.5 to translate user queries into code for visualizations, supporting iterative refinement through multi-turn dialogue. Similarly, pipeline approaches like LIDA~\cite{dibia2023lida} decompose the visualization generation process by combining LLMs with visualization rules. Commercial tools have also adopted LLMs for CVA~\cite{tableauAgent2025,microsoftPowerBIQA2025,googleLookerConversationalAnalytics2024, databricksAIBIGenie2025,thoughtspotSpotter2025}.  BaViSitter~\cite{bavisitter} further expands the CVA landscape by exploring multimodal interfaces, enabling users to issue commands that incorporate both natural language and interactive visual references. Our work complements this by focusing on how to systematically evaluate such interactions, especially when they involve ambiguity, context carryover, and inference.

While these tools demonstrate the potential of LLMs to make data exploration more accessible, questions remain about how practitioners actually use CVA tools in real-world settings and how they assess the quality of generated outputs. In this work, we conduct formative studies to explore practitioners' real-world use of CVA tools, uncovering their evaluation criteria and workflows when assessing CVA tool outputs. Building on these practitioner insights, we present a toolkit that operationalizes these findings into test cases, metrics, and evaluation interfaces to help systematically evaluate, compare, and improve the models and prompts used in LLM-based CVA tools before deployment.


\subsection{CVA Evaluation Methods}
\subsubsection{Benchmarks}
Standardized benchmarks (e.g., BIG-Bench~\cite{srivastava2023beyond}, HELM~\cite{Liang2022HolisticEO}) help evaluate LLMs and system prompts at scale, by specifying inputs and expected outputs across diverse use cases, from reasoning puzzles to even basic analytics questions, making benchmarks central to progress in model evaluation. Data-centric benchmarks, like Spider~\cite{yu2018spider, lei2024spider} and CoSQL~\cite{yu2019cosql}, evaluate natural language to SQL generation. For natural language to visualization generation, the nvBench~\cite{luo2021nvbench}, nvBench 2.0~\cite{luo2025nvbench2} and VisEval~\cite{viseval} test suites provide large-scale natural language to Vega-Lite mappings. However, these benchmarks have important limitations. First, they are synthetically generated rather than derived from real end-user usage, meaning they do not reflect how practitioners actually interact with CVA tools in practice. For instance, most benchmarks focus on single-turn queries, overlooking the multi-turn conversational dynamics (such as context carryover, iterative refinement, and evolving analytical goals) that characterize real-world data exploration. Second, running and adapting these benchmarks requires programming expertise, computational resources, and technical setup (e.g., configuring databases, managing API calls, writing evaluation scripts), and time, creating barriers for practitioners, such as product managers, tool designers, and other low-code stakeholders, who want to evaluate CVA systems without extensive technical overhead.

Our work addresses these gaps through formative studies that observe and log real-world CVA usage by end-user analysts. From this empirical foundation, we derive \lexara's test case suite, which captures authentic usage patterns including composite questions, ambiguous intents, and multi-turn dynamics like context carryover during iterative analysis. Furthermore, \lexara's interface is designed to democratize CVA evaluation: practitioners can upload their own test cases, configure evaluation experiments, and explore results at multiple levels of granularity—all without writing code. This low-code approach makes systematic evaluation and experimentation accessible to a broader range of users, supporting iterative improvement of CVA tools before deployment.

\subsubsection{Interactive Benchmarking Tools}
A growing ecosystem of evaluation tools helps developers probe and debug model behavior. OpenAI Evals, Google's AutoSxS, ChainForge~\cite{Arawjo2024ChainForge}, EvalLM~\cite{Kim2024EvalLM}, and LLM Comparator~\cite{Kahng2024LLMComparatorTVCG} support test case comparison, hypothesis testing, and automated judgment. Other VA tools like PromptIDE~\cite{Strobelt2023PromptIDE}, LMdiff~\cite{Strobelt2021LMdiff}, and Sequence Salience~\cite{Tenney2024SequenceSalience} assist with prompt iteration and token-level inspection. To support model evaluation and debugging, tools such as OpenAI's Evals framework~\cite{openai_getting_started_with_evals_2024} and Google Vertex AI AutoSxS~\cite{google_run_autosxs_pairwise_eval} offer capabilities like side-by-side output comparison, rule-based and LLM-based judging, and hypothesis testing. However, these tools are built to evaluate single-turn, single-format outputs, usually text. 

\lexara~builds on these tools to offer native support for CVA-specific aspects like multi-turn analyses, comparing rendered visualizations, visualization grammars, analytical thinking in natural language responses, based on datasource analysis. Its design incorporates domain-relevant ontology (e.g., ambiguity type, axis match, interactivity) and conversation-aware diagnostics, enabling more targeted debugging and benchmarking of CVA system behavior.

\subsubsection{Human and Automated Evaluation Methods}
Given the complexity of CVA interactions, human evaluation remains the gold standard. Experts or users must manually assess nuances and rate responses on a range of criteria. However, this process is labor- and time-intensive, and does not scale well across large prompt sets and model configurations~\cite{belz-etal-2023-non,clark2021all,howcroft-etal-2020-twenty}. To scale evaluation beyond manual methods, LLM-as-a-Judge approaches are increasingly used to assess model outputs along dimensions like coherence, correctness, or explanation quality~\cite{ni2024systematic,gu2024survey,kim2024prometheus}. These methods often correlate better with human judgment than traditional metrics, particularly in open-ended QA, vision-language tasks, and agent reasoning~\cite{li2023generative,Chang2024LLMEvalSurvey}. Yet, studies show systemic biases — including self-preference, verbosity, position, and concreteness biases~\cite{Zheng2023MTBench,wataoka2024self,shi2024judging,ye2024justice}, which can skew results if not addressed. \lexara~ develops a complementary hybrid human-AI evaluation approach to balance the richness of user-centered values, evaluation criteria and methods, with the scale of LLM-as-a-Judge automated methods. To ensure validity and reliability of evaluation methods, \lexara implements prompt, model, and interface–level safeguards: Evaluation criteria are derived from a formative study with practitioners and end-users. It employs few-shot prompts seeded with end-user–labeled evaluation examples that reflect analyst values and highlight common points of confusion. The LLM-as-a-Judge Recommendation feature recommends models from outside the candidate model family to reduce self-preference, and evaluation runs randomize item positioning and use per-output scoring against a reference rather than pairwise comparisons to limit position bias. Prompts explicitly instruct judges to ignore stylistic flourish and avoid rewarding verbosity by truncating or equalizing answer length. To counter concreteness and stylistic biases, \lexara~uses detailed rubrics that specify grounded analytical criteria and provides end-user-annotated examples across multiple orthogonal metrics. The interface further surfaces judge rationales, JavaScript Object Notation (JSON) spec diffs, and rendered charts, enabling human inspection and override in a hybrid workflow. Finally, we validate \lexara's metrics against human raters (e.g., Cohen’s $\kappa$, Spearman $\rho$) to ensure alignment.

\subsection{Evaluation Metrics for Visualization and Analytical Language}
Evaluating visualizations requires balancing correctness with interpretability and usability. Foundational work has emphasized the trade-offs between ecological validity and experimental rigor~\cite{plaisant2004challenge,zuk:2006}, advocating for mixed methods~\cite{lam2011empirical,islam:2024} and cognition-grounded metrics~\cite{Carpendale2008,battle:2018,liiv:2010}. More recently, automatic evaluation techniques have emerged. VIS-Shepherd~\cite{VISShepherd2025} employs multimodal or LLM critics to rate visualization quality. Vi(E)va LLM! \cite{podo2024vievallmconceptualstack} proposes a layered stack—from code similarity to insightfulness—applying measures like Jaccard similarity, SSIM, and VLAT. SimVecVis~\cite{liu2025simvecvisdatasetenhancingmllms} encodes chart structure as latent vectors and evaluates reconstruction performance. Song et al.~\cite{song:2024} raise critical methodological questions about evaluating LLM-generated visualizations, arguing traditional metrics fail to capture visual design's complexity and subjectivity. They advocate for nuanced, design-aware strategies considering interpretability, expressiveness, and task relevance -- aligning with Lexara's graded, hybrid approach.

In parallel, natural language evaluation has matured from n-gram metrics (e.g., F1, Precision, Recall~\cite{10.1145/3606367}, BLEU~\cite{papineni2002bleu}, ROUGE~\cite{lin2004rouge}) to semantic measures (e.g., BERTScore~\cite{zhang2019bertscore}, BLEURT~\cite{sellam2020bleurt}). Yet, these often fail to capture reasoning quality, contextual coherence, or domain-specific accuracy, especially in open-ended CVA explanations. Addressing these limitations, \lexara develops hybrid visualization and language metrics that accommodate ambiguity, support multiple correct outputs, and integrate rule-based and rubric-guided LLM-as-a-Judge pipelines. Users can inspect, override, and refine judgments, enabling interpretable, scalable automation with human-in-the-loop evaluation.

\section{Formative Studies: Eliciting Real-World Use Cases, Evaluation Criteria \& Workflows}
\label{sec:formative}
While prior tools and benchmarks have advanced model evaluation, they often overlook the actual experiences of CVA practitioners, i.e., those building or using these tools in real-world settings. To ground the design of the \lexara~toolkit in these practitioner perspectives, we conducted two complementary formative studies: interviews with tool developers to surface design rationales and evaluation workflows, and observational sessions with end-users to capture situated judgment during CVA scenarios. Taken together, these complementary approaches revealed both the considerations shaping system design and the situated practices of use, grounding our toolkit in the perspectives of those who build and those who rely on CVA tools.

\subsection{Study 1: Tool Developers' Use Cases, Evaluation Criteria \& Workflows}
We conducted one-hour semi-structured video interviews with 22 professionals involved in developing CVA tools. Participants included researchers, designers, engineers, and product managers. Using a snowball sampling strategy~\cite{atkinson2001accessing,biernacki1981snowball,noy2008snowball}: we initially reached out to subject matter experts in CVA and LLM evaluation in a large technology company, who then recommended additional colleagues with relevant expertise. This sampling ensured coverage across product, research, and engineering roles engaged in building or evaluating LLM-powered CVA systems. The interviews explored four key areas: (1) use cases of CVA tools, (2) evaluation criteria, (3) workflows used to assess models and system prompts for these use cases, and (4) any challenges they faced in conducting these evaluations. All participants gave informed consent for data collection (audio, video, and usage logs). Sessions were recorded, transcribed, and thematically analyzed using an open-coding approach \cite{charmaz2014constructing} to identify use cases, evaluation workflows and their challenges, and evaluation criteria. Refer to the supplementary materials for details on our experimental protocol, interview guides and apparatus.

\subsection{Study 2: End-Users' Use Cases, Evaluation Criteria \& Workflows}
We conducted 45-minute lab-based sessions with 16 professional data analysts or end-users ($U1$-$U16$) across diverse domains including finance, education, healthcare, and technology. Participants held roles such as analysts, BI advisors, data architects, research scientists, consultants, and product managers. Recruitment was conducted via a visual analytics conference, supplemented by direct outreach to attendees interested in conversational AI and visualization tools. This strategy provided access to participants who had real-world VA experience but varying familiarity with conversational interfaces (six beginners, seven intermediate, three advanced). We classified participants as beginners if they reported less than one year of regular experience using conversational interfaces for data analysis. Intermediate participants reported approximately 1–3 years of experience with BI tools and occasional authoring of visualizations. Advanced participants had more than three years of experience building or maintaining analytics workflows and routinely authored or reviewed visualizations. Each session had two phases:

\pheading{Phase 1: Think-Aloud CVA Interaction [15–20 min]}
To understand their usage of CVA tools, participants used a commonly-used commercial LLM-enabled CVA tool \cite{tableauAgent2025} to analyze a datasource of their choice, either from a curated gallery or their own  (Appendix Table \ref{tab:formative-datasets}). A Chrome extension (Appendix Figure \ref{fig:tc-extension}) recorded their multi-turn interactions, capturing prompts, model responses, and in-the-moment reflections. After each response, participants rated its quality using Likert-style criteria and corrected outputs when needed to reflect their expectations. Participants could also suggest custom evaluation criteria or flag inaccuracies. These real-world logs later informed the design of \lexara's test case library (see Supplementary Materials).

\pheading{Phase 2: Side-by-Side LLM Response Comparison [20–25 min]} Participants compared anonymized outputs from multiple models (GPT-4o, Claude-Opus-4, GPT-o3 anonymized for participants to avoid biasing) for the same user utterances (visualizations, natural language responses, and JSON grammar specifications, alongside traditional metrics (F1, Precision, Recall \cite{10.1145/3606367}) (Appendix Figure \ref{fig:model-diffs}). Grammar specifications were shown to expose structural decisions such as field encodings, chart types, filters, and sort logic -- details not always visible in the rendered visualizations. By surfacing specifications alongside outputs, we enabled participants to diagnose why two similar-looking charts diverged and express expectations around cross-format consistency~\cite{setlur2016eviza,tory2019mean}.

They were asked to think aloud as they compared model outputs, evaluation trade-offs, and reflected on where existing metrics fell short. All sessions worked with the same utterances from the Superstore datasource~\cite{tableauSuperstore}, which is part of the NLVCorpus benchmark~\cite{srinivasan2021collecting}. We selected Superstore due to its familiar business context and rich analytical scope, which could elicit ambiguity, multi-turn reasoning, and diverse chart types~\cite{setlur2016eviza,setlur2022you}. Using this shared datasource ensured ecological validity while enabling consistent comparisons across sessions.

\subsection{Characteristics of CVA Use Cases}
In addition to analyzing developer and end-user interviews, we thematically analyzed utterances from Phase 1 of Formative Study 2 (\S\ref{sec:formative}.2), where 16 data professionals from domains, such as finance, education, healthcare, and technology engaged in multi-turn CVA sessions ($\sum=$ 80 utterances, $\mu=5.8, \sigma=3.1$ turns per conversation). Using a browser extension (Appendix Figure \ref{fig:tc-extension}), we logged user–system interactions, including in-the-loop ratings, corrections, and labels for each utterance. The resulting conversations spanned a diverse set of utterance types and evaluation challenges, reflecting realistic task complexity and variation across domains (finance (8 conversations), education (3), healthcare (5)). This annotated set of utterances reflects the diversity of analytical intents and reveals key challenges in interpreting user intent during conversational interactions, described as follows:

\subsubsection{Visualization Types}
Thematic analysis of $64$ user utterances revealed requests for a diverse range of chart types: bar chart (n=30), scatter plot (n=6), line chart (n=18), box plot (n=4), histogram (n=3), multi-line chart (n=3). For example, for the utterance, \textit{``Plot a scatter of \texttt{discount} vs. \texttt{profit margin},''}, $U5$ remarked, \textit{``I expected a scatter plot, but it gave me a bar chart. Technically valid but not what I asked''}. Such examples underscore the importance of semantic precision and visual format alignment in user expectations, highlighting the need for evaluation metrics sensitive to visualization intent and not just syntactic correctness.

\subsubsection{Ambiguity in User Utterances} 
In analyzing the user utterances, we observed that $27$ utterances exhibited some form of ambiguity, requiring the system to make context-sensitive inferences. Ambiguity in natural language, defined as the presence of multiple plausible interpretations for a single expression or request, is a well-documented challenge in CVA interfaces~\cite{Alharbi:2012,gao2015datatone,setlur2016eviza}. Ambiguity often arises when user intent is underspecified, field references are vague or mismatched, or contextual cues from earlier turns are required for correct interpretation. A single utterance could display multiple forms of ambiguity.

\pheading{Syntactic Ambiguity.} 18 utterances demonstrated syntactic ambiguity, which arises when the structure of a sentence permits multiple grammatical interpretations~\cite{Jurafskybook}. For example, $U13$ asked, \textit{``Show top 10 products in furniture by sales region with high profit,''}, and responded to the result: \textit{``I could see at least two ways to interpret that,} [Show the top 10 products in the furniture category, grouped by sales region, but only include those with high profit. or Show the top 10 products in the furniture category, by those sales regions that have high profit.] \textit{and the model picked one} [the latter].'' 

\pheading{Semantic Ambiguity.} Of the $27$ ambiguous utterances, $19$ involved semantic ambiguity, wherein an utterance could plausibly map to multiple fields or concepts in the datasource, due to underspecified or imprecise language. For example, $U4$ asked, \textit{``Show me profit over time''}, although the datasource only contained fields  \texttt{Net Revenue} and \texttt{Cost}. The model inferred a plausible mapping and returned a chart plotting Net Revenue over time. The user later reflected, \textit{``I said `profit,' but in this datasource that could mean revenue minus cost, or maybe net sales. The model guessed \texttt{Net Revenue}.}'' 

\pheading{Pragmatic Ambiguity.} Pragmatic ambiguity arises when the meaning of a user’s prompt depends on context, whether from earlier dialogue turns or implicit assumptions that are not explicitly stated in the input utterance~\cite{setlur2016eviza,tory2019mean}. Participants encountered such ambiguity in $37$ utterances, and judged systems not only by their final outputs, but by how well they interpreted underspecified prompts in light of surrounding context. 

Participants often incrementally elaborated on previous prompts ($19$ utterances), expecting earlier filters or visual structure to persist. For example, $U7$ began with, \textit{``Show sales by region''}, followed by, \textit{``Now break it down by category.''} They appreciated when the model preserved the region filter, noting,  \textit{``I like that it remembered my earlier region filter, but sometimes other models just dropped it.''}

Ambiguity also arose when users referred back to previously mentioned entities without explicitly naming them again ($32$ utterances). For example, when $U12$ asked, \textit{``Which of these categories had the highest growth?''} (following a turn which asked about product categories) \textit{``It got confused about what 'these categories' meant and pulled in something else [the model erroneously filtered out all the categories returning a vacuous chart.].''} Participants also expected temporal or categorical filters to persist across related turns. $U2$ explained how this shaped their interpretation when they asked, \textit{``Show only \texttt{2023 sales}''} next \textit{``Now compare \texttt{East} vs. \texttt{West}.''} \textit{``When the filter carried through, the analysis made sense. When it didn't, it felt like it didn't get me and I had to keep starting over.''}

Another form of pragmatic ambiguity involved both implied concepts and underspecified utterances ($64$ occurrences), where users referenced fields, filters, sorts, or time units without explicitly naming them, expecting the system to resolve intent through context. For example, $U5$ implicitly assumed descending order for a visualization response to the utterance, \textit{``Top 10 states by profit.''} They remarked, \textit{``It didn't sort descending, so the top 10 wasn't actually top.''}

\subsection{Evaluation Criteria for CVA Use Cases}
\label{sec:criteria}
These utterances provide a foundation for systematic evaluation. Building on this foundation, we examined how practitioners actually judge the quality of CVA outputs in real-world scenarios. Through thematic analysis of the data, we found that participants consistently evaluated responses along three key categories: \textit{Visualization Quality}, \textit{Natural Language Quality}, and \textit{Conversation Quality}. 

\subsubsection{Visualization Quality}
\begin{figure*}[t!]
\centering
\includegraphics[width=\textwidth]{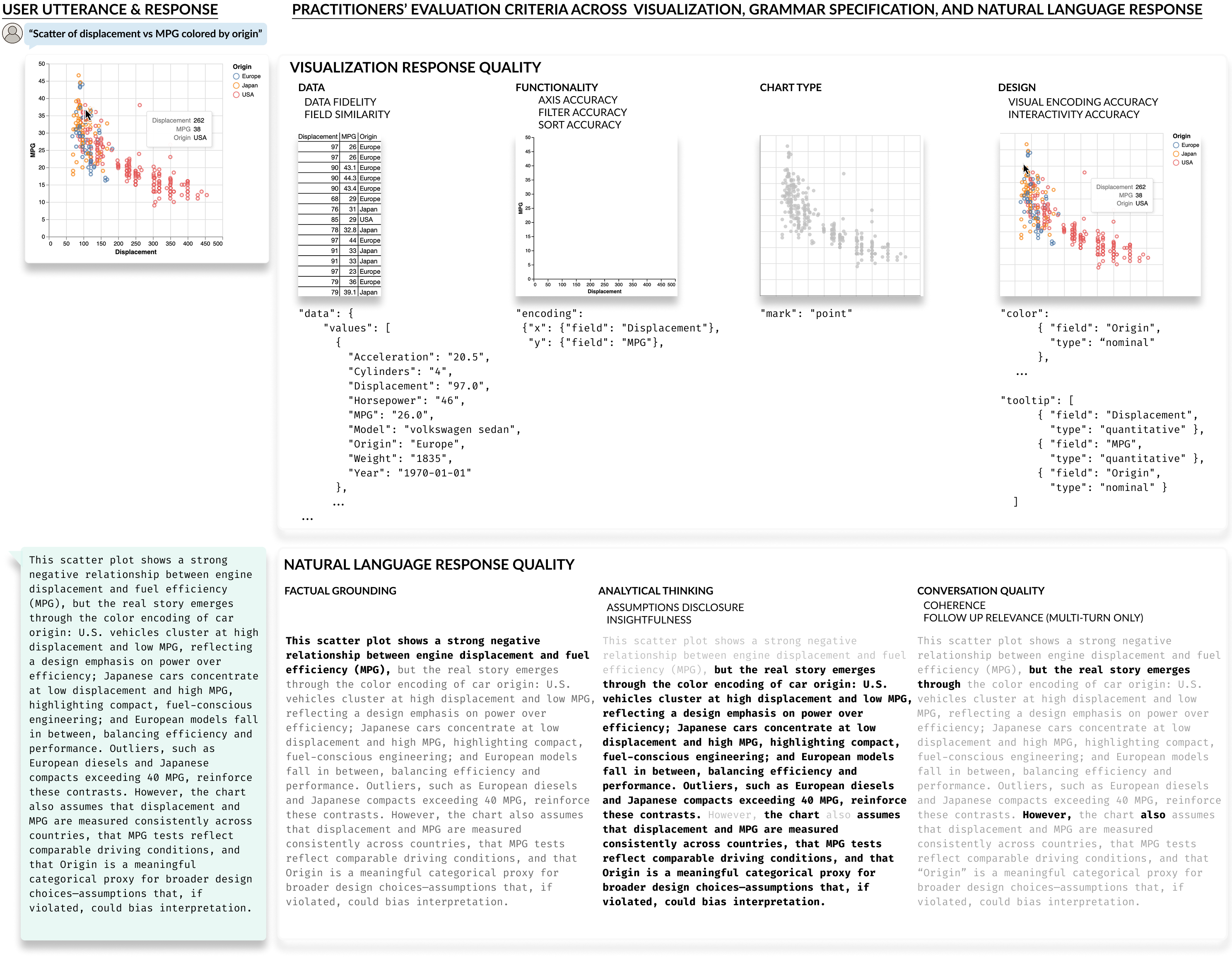}
\caption{Illustration of how practitioners evaluate the multi-format CVA response. The example shows (left) a user utterance and corresponding model outputs, and (right) the evaluation criteria identified in our formative studies: visualization response quality assessed by looking at both the rendered visualization and grammar specification (data fidelity, chart type, functionality, design), natural language response quality (factual grounding, analytical thinking and conversation quality). This figure provides a conceptual overview and does not reflect the actual UI of any CVA system.}
\label{fig:setupEval}
\Description{This figure shows how a user’s request, “Scatter of displacement vs MPG colored by origin,” is evaluated. At the top left, the raw data table is partially shown, with fields such as acceleration, cylinders, displacement, horsepower, and origin. To its right are partial JSON snippets specifying encodings: displacement on the x-axis and MPG on the y-axis, with “mark: point.” Further to the right, scatterplots illustrate different evaluation dimensions. The first scatterplot shows only functionality, plotting displacement versus MPG. The next highlights chart type as a scatterplot, followed by a version adding design with color encoding for car origin (Europe, USA, Japan). The rightmost scatterplot shows the final visualization with all elements together. Below the scatterplots, four blocks of natural language evaluation criteria are presented: “Factual Grounding” describes the negative relationship between engine displacement and MPG. “Insightfulness” highlights how different origins cluster by efficiency and performance. “Assumptions Disclosure” notes assumptions about consistent measurement across countries. “Conversation Quality” combines these insights into a coherent explanation. Together, the figure demonstrates how evaluators assess visualization outputs and language responses across multiple dimensions.}
\end{figure*} 

Visualization Quality refers to how accurately and appropriately a generated visualization represents the correctness of data values, the appropriateness of chart types for the given analytical intent, and the presence of filters and sorting operations applied to the underlying data representing the visualization. Based on participant responses, we group visualization quality concerns into four key categories: \textit{Data}, \textit{Chart Type}, \textit{Functionality}, and \textit{Design}.

\pheading{Data.} We begin with criteria related to the data underpinning the visualization, specifically, whether the data is faithfully represented (\textit{Data Fidelity}) and whether the fields selected align semantically with user intent, even when exact matches are absent (\textit{Field Similarity}). 

Participants emphasized that charts must truthfully and completely reflect the underlying data; any deviation, such as missing, extra, or incorrectly aggregated rows or columns, was viewed as a major breach of trust (Data Fidelity). As $U11$ points out, \textit{``If it says Profit but it’s clearly counting rows, that’s a fail for me.''}. Participants allowed partial credit when the underlying data was what they expected, but the analytical operation applied to the data (e.g., aggregation type) was not what they expected. For instance, $U2$ remarked, when viewing `count' as the aggregation type rather than `sum': \textit{``It's not a big deal but here it seems to have missed \texttt{Sum of Quantity}''} ).

Another critical dimension of data correctness is whether the fields appropriately reflect the user's analytical intent. This involves both syntactic matching (e.g., field names) and semantic understanding of the utterance (Field Similarity). Participants emphasized that fields bound to visual elements must align with the task intent, even if the exact names do not match. Near‑misses, such as mapping a user's mention of ``sales'' to an attribute labeled \texttt{sales\_amount}, or interpreting ``date'' as \texttt{Billing\_Timestamp\_HFD}, were typically accepted when types were compatible or meanings were semantically close. As $U14$ explained, \textit{``in enterprise schemas, fields rarely match user wording exactly and reasonable inferences preserve flow.''} Participants also appreciated when the model could make these inferences but strongly preferred transparency in the mappings. $U14$ continued, \textit{``Recognize \texttt{date} even when the column is \texttt{Billing\_Timestamp\_HFD}, but tell me what you picked. Bold the exact field names you chose. ''}

\pheading{Chart Type.} Participants consistently emphasized that the appropriateness of the chart type directly impacts how easily they can interpret and trust the visual output. They expected models to follow established visualization best practices~\cite{showme}, such as generating line charts for trends over time or bar charts for categorical comparisons. For instance, $U8$ noted upon seeing a bar chart being generated, \textit{``Two models picked the wrong chart for profit per month: it should be a line. I'll still give them partial credit, but they didn't pick the best chart.''}

\pheading{Functionality.} Participants assessed functional correctness based on whether the visualization functioned as expected in terms of axes, filters, and sorting choices. Specifically, participants expected the axes to reflect correct field mappings, orientation, and scale, including the use of zero baselines when appropriate. $U15$ remarked, \textit{``This one doesn't start at zero. That’s misleading''} , and \textit{``Can we get units on the axis?''} Swapped axes (e.g., X and Y reversed) were generally not treated as outright failures, since the data geometry remained valid, but participants felt they deserved a moderate score rather than full credit. However, critical errors like incorrect scale type or missing baselines, were scored lower. \textit{``Missing titles or units are not deal-breakers, but more like nice-to-haves''} ($U14$).
      
Participants viewed filter correctness as essential for analytic continuity. They expected the models to clearly indicate applied filters and penalized both over-filtering and under-filtering.  $U3$ and $U5$ correspondingly stated, \textit{``Only this model didn't filter. That's the first thing I check''} and \textit{``This one added extra \texttt{Year=2024}.''} 
        
Sorting accuracy was similarly important for data prioritization tasks. Participants expected the sorting behavior to match either explicit instructions or be reasonably inferred from context. When reviewing the model output for \textit{``Show top 10 products in Furniture by Sales''}, $U5$ commented, \textit{``All the other models chose to \texttt{Sales} descending, but seems like this model chose to not do that. I guess it's ok because it's implied but not expected.''}

\pheading{Design.} Participants assessed design quality based on how truthfully and clearly the visualization encoded information, prioritizing functionality and interpretability over any stylistic enhancements.

In particular, participants expected visual encodings, such as color, size, shape, opacity, and text labels to reflect meaningful distinctions in the data. Participants attempted to refine visual encodings through follow-up prompting:  \textit{``Color by Region; add data labels.''} ($U10$). 

Participants also expected interactive affordances, particularly tooltips to reveal accurate, relevant data on demand. When these elements were incorrect or incomplete, it broke their flow and raised concerns about the model's reliability. $U8$ noted, \textit{``I hovered and it showed the wrong value. Tooltip said `sum' but it was a count.''}  Participants, such as $U11$ also prompted to explicitly surface additional information - \textit{``Include \texttt{Sales} and \texttt{Profit} in the tooltip.''}

\subsubsection{Natural‑Language Response Quality}
In addition to visualization correctness, participants carefully evaluated the accompanying natural language responses, particularly when models provided textual explanations or summaries alongside charts.

\pheading{Factual Grounding.} Participants consistently prioritized factual consistency between the chart and the text. They expected that descriptions include all salient facts, such as filters, measures, magnitudes, and directional trends that were encoded visually. If key facts were missing or incorrect, trust was quickly eroded and treated as high-severity errors. As $U12$ emphasized, \textit{``If the chart and the text disagree, I stop trusting either.''}

\pheading{Analytical Thinking.} Participants also evaluated how well the system reasoned alongside them, looking for evidence of reasoning or interpretation beyond simple description. In particular, participants appreciated when the model explicitly surfaced filters, timeframes, or aggregation logic, i.e., assumptions disclosure. This transparency helped them understand and verify how results were derived. For instance, $U11$ noted, \textit{``If you assume individual profit values, say so.''} 

Higher-rated responses that synthesized trends, pointed out anomalies, and suggested comparisons or causes, were considered insightful; rather than restating input queries, participants valued responses that proactively explained what the data meant. As $U8$ shared, \textit{``This one points out that sales dropped in Q4. That's helpful''}  and \textit{``I would like to get to a point where these systems just give me rich actionable insights not just say I did what you asked me to''} ($U11$).

\subsubsection{Conversation Quality}
Beyond isolated responses, participants evaluated how well CVA systems sustained coherent, context-aware conversations over multiple turns. This process included judging whether the system maintained logical flow, preserved contextual intent across prompts, and adapted outputs based on evolving dialogue.

\pheading{Coherence.} Participants valued responses that were logically structured, internally consistent, and free from contradictions. They often praised outputs that maintained a clear reasoning chain and articulated how different observations connected. For example, $U3$ said, 
\textit{``What its saying makes sense that sales rose in Q4 so inventory dropped. So, this could impact next quarter.''}

 \pheading{Follow‑up Relevance.} Participants emphasized that in multi‑turn interactions, the model must retain prior context: including applied filters, selected categories, or inferred user goals. Outputs that failed to carry over context felt disjointed or inattentive. $U6$ highlighted for this utterance, \textit{``Focus on high-growth segments in Q3 only,''} when the model added a filter with Q3's dates instead of the whole year, they commented, \textit{``I like that since we asked about high-growth segments in Q3, this tells me what happened in Q3 only.''} In \lexara, we operationalize these concerns by scoring each response in situ with respect to its preceding conversational context, rather than collapsing an entire dialogue into a single scalar score. This per-turn, context-aware design lets practitioners see where multi-turn workflows recover from errors or break down.

\subsection{Workflow Challenges and Design Considerations}
\label{sec:workflows-challenges}
Through interviews with the CVA practitioners, we identified five core challenges (C1–C5) that they face when evaluating CVA systems. Each challenge is associated with recurring evaluation workflows, which, while common, often fall short of supporting systematic and scalable LLM benchmark evaluation.

\pheading{C1: Fragmented, ad hoc comparisons.} Practitioners primarily relied on manual side-by-side comparisons of models and prompts. They often tested the same utterance across configurations, consolidated outputs in spreadsheets or slides, and visually inspected screenshots and specs. $T3$ and $T11$ explain their respective workflows, \textit{``I literally had tabs for each model. One with the spec, one with screenshots, and then I'd eyeball which chart dropped categories''}  and \textit{``I feel like I'm always just context switching across all these channels, which leaves me not able to have time for really diving into types of behavior I care about.''}

\pheading{C2: Misalignment with domain-specific tasks.} While public benchmarks like nvBench~\cite{luo2021nvbench}, Spider~\cite{lei2024spider}, and VisEval~\cite{viseval} offered a shared vocabulary, practitioners found them poorly aligned with their specific CVA use cases. They often needed to test ambiguous field references, vague temporal phrases, or domain-specific analytic tasks that benchmarks did not capture. $T17$ stated, \textit{``Aggregate scores rarely tell me if the model will mess up axes when I ask for 'profit by segment' and such''} and $U6$ expressed frustration when describing their workflow: \textit{``need to tailor to own data and use cases.''} 

\pheading{C3: Unreliable transfer to actual environments.} Practitioners frequently encountered mismatches between public benchmark performance and real-world reliability. Models that performed well in papers or demos often broke down when applied to internal datasources or production workflows. As $T20$ noted, \textit{``Prompts work for their demo datasource, but failed on ours.''} To get more reliable comparisons, some developers used programmatic test suites, running benchmark scripts over curated utterances in notebooks. While this added reproducibility, the gap between test coverage and domain-specific needs remained. $T7$ explains, \textit{``We run our benchmark on real data and look for accuracy, cost, speed columns, and export machine-readable logs.''}

\pheading{C4: Inaccessible and opaque evaluation outputs.} Evaluation pipelines often required programming expertise and produced outputs that are typically JSON logs or console traces. This limited collaboration between engineers, PMs, and designers from participating meaningfully in the evaluation, as $T17$ points out: \textit{``We run programmatic scripts from nvBench, but the outputs are JSON logs that PMs can’t interpret.''}  Even for technical users, understanding \emph{why} a case passed or failed remained difficult as few tools supported granular inspection or comparison across meaningful categories - \textit{``No self-service so the benchmarking evaluation dashboard is hard to update and maintain, so results stay opaque to non-engineers''} ($T7$). 

\pheading{C5: Lack of interpretable and graded metrics.} Participants reported that standard metrics, such as accuracy, BLEU, F1 rarely captured the graded, nuanced correctness required in CVA. Many cases involved partial correctness (e.g., correct chart but misleading axis), where binary scoring fell short, as $T14$ pointed out, \textit{``In our work, accuracy isn't just yes or no. Sometimes it's close enough to be useful, other times a valid looking chart is misleading.''}

\section{Design Considerations For A CVA Evaluation Toolkit} 
\label{sec:designconsiderations}
Guided by the evaluation workflows and challenges (C1–C5) outlined in \S\ref{sec:workflows-challenges}, we define seven design goals (D1–D7) to guide the evaluation of an LLM-based CVA toolkit (i.e., a software system that integrates reusable components for building, testing, and analyzing CVA test cases). Each goal is grounded in observed needs and translated into concrete design strategies, which we realize in the \lexara~toolkit (\S\ref{sec:lexara}).

\pheading{D1: Lower the barrier to systematic benchmarking.} Enable low-code practitioners to set up, run, and interpret benchmarking experiments with minimal effort based on challenges \textbf{[C1]} and \textbf{[C4]}. By providing templates for datasources, utterance sets, rubrics, and sensible defaults for metrics and comparisons, the toolkit should support reproducible evaluations without requiring programming skills, a need emphasized in HCI literature~\cite{ribeiro-etal-2020-beyond,kiela-etal-2021-dynabench}.

\pheading{D2: Tailor evaluations to real-world CVA use cases.} To address challenges of misalignment with real-world CVA tasks (\textbf{[C2]}, \textbf{[C3]}), the toolkit should support benchmarking on user-specific data, tasks, and prompts, a paradigm advocated for contextualized LLM evaluation aligned with practitioners' tasks and goals~\cite{gardner-etal-2020-evaluating}.

\pheading{D3: Scale evaluations with speed and reliability.} To address the bottlenecks in manual comparison (\textbf{[C1]}) and opaque tooling (\textbf{[C4]}), the toolkit should support scalable, repeatable experiments across many utterances, prompts, and models, enabling efficient comparisons and rapid iteration~\cite{dodge-etal-2019-show}.

\pheading{D4: Compare across formats.} To address challenges in manual, fragmented comparisons (\textbf{[C1]}), the evaluation toolkit should support reasoning on alignment across multiple output formats (i.e., rendered visualizations, natural language explanations, and underlying chart specifications)~\cite{masry-etal-2022-chartqa}.

\pheading{D5: Link overviews to instance-level insights.} To address limitations in fragmented comparisons and opaque tooling (\textbf{[C1]}, \textbf{[C4]}), the toolkit should allow practitioners to fluidly navigate between high-level aggregate metrics and fine-grained utterance-level results. This supports practitioners in identifying performance subsets across task types, semantic categories, or failure modes that are most relevant to their analysis goals~\cite{lunardi2025robustnessreliabilitybenchmarkbasedevaluation}.

\pheading{D6: Support context-aware diagnostic analysis.} Evaluation toolkits should enable practitioners to interpret model behavior in relation to specific analytic contexts (\textbf{[C1]}, \textbf{[C4]}). Drawing from prior work in model interpretability~\cite{ribeiro-etal-2020-beyond,dodge-etal-2019-show}, the toolkit should identify when models succeed or fail (e.g., across utterance types, data domains, or prompt strategies), and how those outcomes emerge through qualitative and quantitative patterns in model and prompt behavior.

\pheading{D7: Make metrics interpretable and actionable for handling multiple plausible answers.} To address limitations in binary metrics (\textbf{[C5]}), CVA evaluations should employ transparent, graded metrics that can accommodate multiple valid outputs. Metrics should clearly communicate what is being measured, why it matters, and how the score was derived to support trust and actionability for model and prompt selection~\cite{gardner-etal-2020-evaluating,kiela-etal-2021-dynabench}.

\section{\lexara: A User-Centered CVA Evaluation Toolkit} 
\label{sec:lexara} 
Building on the design considerations, we developed \lexara, a user-centered evaluation toolkit that operationalizes the findings from our formative studies. The toolkit comprises three complementary components:

\begin{tightitemize}
    \item \textbf{Test Cases from Real-world CVA Conversations} [D2, D6] [\S\ref{sec:testcases}]: A curated suite of multi-turn user queries annotated with expected outputs and labeled for prevalent CVA challenges such as ambiguity, inferred fields, and context carryover.
    
    \item \textbf{CVA Evaluation Metrics} [D7][\S\ref{sec:metrics}]: A set of interpretable metrics derived from practitioners' evaluation criteria covering visualization quality, analytical natural language response quality, and conversational coherence, designed to handle graded correctness and multiple plausible answers.
   
    \item \textbf{CVA-Specific Interactive Evaluation Tool} [D1, D3, D4-6]  [\S\ref{sec:tool}]: A low-code interface for setting up and running evaluations, comparing multi-format outputs across models, prompts, and turns, and linking aggregate metrics to fine-grained diagnostics based on structured test case templates.
\end{tightitemize}

\subsection{\lexara's Test Cases from Real-world CVA Conversations}
\label{sec:testcases}
Each test case is based on a specific datasource and is represented in YAML/JSON format. The datasource file includes the following fields: \texttt{title}, \texttt{data-source-fields} the data attributes, each with a \texttt{name} and \texttt{fieldValues}, a column-vector of data values. Each test case includes: \texttt{Conversation ID}, set of \texttt{Utterances} with canonical phrasing (i.e., a representative or standard formulation of the intent) and participant-authored variations, \texttt{Labels} denoting chart type, ambiguity type, context-handling, and inferencing types. The \texttt{expected response} is provided in two formats: (i) a visualization specification (in a JSON schema similar to Vega-Lite~\cite{vega_specification}, covering fields, encodings, transforms, filters, and sorts) and (ii) a natural language explanation that faithfully describes the chart while surfacing assumptions made by the model (e.g., inferred fields or grouping logic). We expose these templates in the User Interface (UI) to give practitioners fine-grained control over datasources and labels at the cost of some upfront schema familiarity.

To ensure these \texttt{expected responses} reflect real-world practitioner or expert consensus, test cases were sourced either from (i) end-user analyst interactions during the formative study, where participants corrected model outputs during real-world usage, or (ii) from prior literature and benchmarks~\cite{luo2021nvbench, viseval, srinivasan2021collecting}. To ensure quality, each expected response was independently reviewed by two CVA domain experts. In cases of disagreement, a third expert adjudicated the final answer. Inter-rater reliability analysis (Cohen's Kappa = 0.81) confirms a high level of agreement across annotators. Since many tasks often have multiple plausible answers, we mitigate subjectivity by explicitly labeling the sources of ambiguity (i.e., syntactic, semantic, and pragmatic) and supporting multiple acceptable expected answers when justified. \lexara's evaluation framework applies graded metrics that capture partial correctness on a continuum, enabling finer-grained diagnostic feedback beyond binary judgments. Because the metrics compute scores by comparing model outputs against these expected CVA responses, the current toolkit is explicitly designed for curated test suites, which as mentioned in the formative study, many teams already maintain. The full test suite is included in the Supplementary Materials and can be directly uploaded into the \lexara~toolkit for benchmarking. 

\subsection{\lexara's User-Centered Evaluation Metrics}
\label{sec:metrics}
{%
\let\Oldincludegraphics\includegraphics
\renewcommand{\includegraphics}[2][]{%
  \Oldincludegraphics[width=1in,height=1in,keepaspectratio]{#2}%
}
\renewcommand{\arraystretch}{1.15}
\newcolumntype{Y}{>{\RaggedRight\arraybackslash\hspace{0pt}}X}
\newcolumntype{T}{>{\RaggedRight\arraybackslash}p{0.9em}}
\begin{table*}[htbp]
\centering
\scriptsize
\begin{tabularx}{\textwidth}{@{}T Y Y Y Y@{}}
\toprule
\textbf{Test} & \textbf{User Utterance} & \textbf{Expected Response} & \textbf{Model Response} & \textbf{Metrics} \\
\midrule

1 Turn 1 &
\codewrap{\texttt{<Quantity>} on y-axis and \texttt{<Region>} on x-axis} &
\begin{minipage}[t]{\linewidth}\vspace{0pt}\centering
\includegraphics{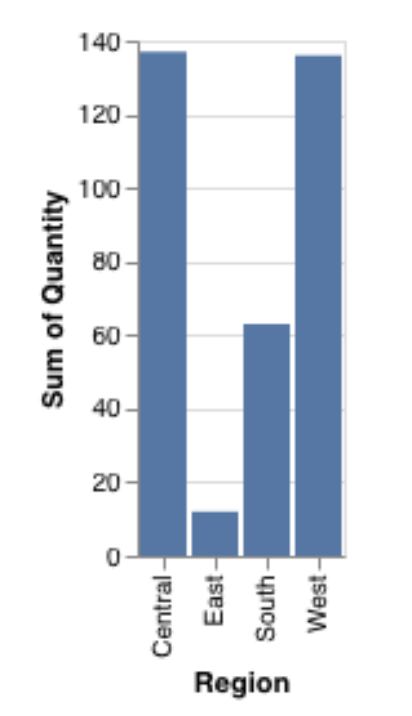}\vspace{2pt}\par
\RaggedRight\footnotesize A vertical bar chart with Region on the x-axis and the sum of Quantity on the y-axis (range 0--140). The Central region has the highest total (~137), with other regions lower.
\end{minipage} &
\begin{minipage}[t]{\linewidth}\vspace{0pt}\centering
\includegraphics{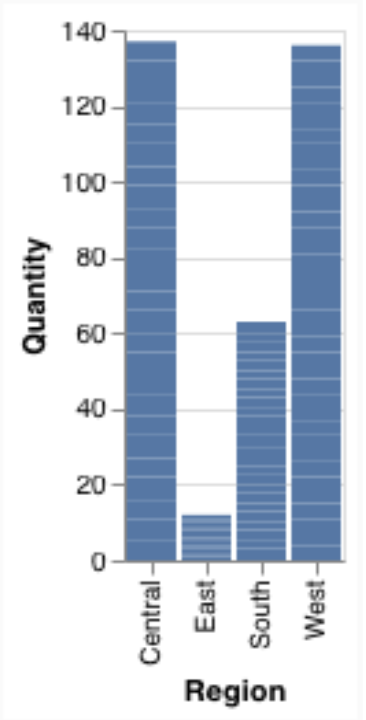}\vspace{2pt}\par
\RaggedRight\footnotesize A bar chart comparing Quantity (y-axis) across Regions (x-axis), allowing a quick scan of how quantities differ by region.
\end{minipage} &
\begin{minipage}[t]{\linewidth}\vspace{0pt}\RaggedRight\footnotesize
Data Fidelity = \scorebadge{100\%}; Field Similarity = \scorebadge{100\%}; Chart Type Similarity = \scorebadge{100\%}; Axis Accuracy = \scorebadge{100\%}; Filter Accuracy = \scorebadge{100\%}; Sort Accuracy = \scorebadge{100\%}; Visual Encoding Accuracy = \scorebadge{100\%}.\\
\textbf{Overall Viz} = \scorebadge{100\%}.\\
NL: Factual Grounding \scorebadge{70\%}; Assumptions Disclosure = \scorebadge{40\%}; Insightfulness = \scorebadge{40\%}; Coherence \scorebadge{90\%}.\\
\textbf{Overall NL} \scorebadge{65\%}.
\end{minipage} \\

1 Turn 2 &
\codewrap{\texttt{Sort by <Quantity>}} &
\begin{minipage}[t]{\linewidth}\vspace{0pt}\centering
\includegraphics{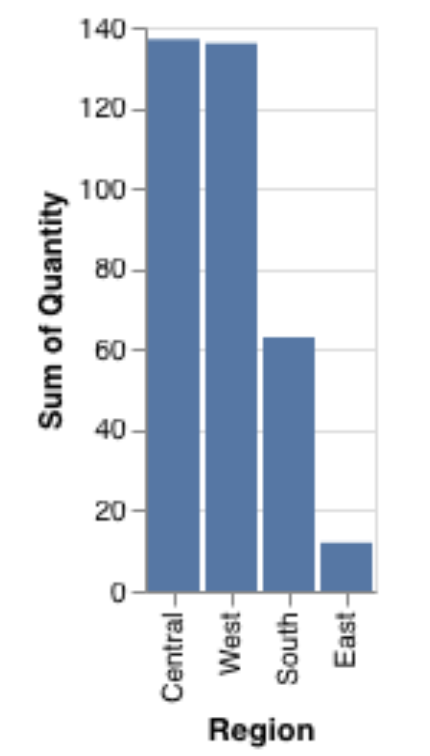}\vspace{2pt}\par
\RaggedRight\footnotesize A vertical bar chart of Regions, sorted in descending order by the sum of Quantity, so the highest-quantity region (Central, ~137) appears first.
\end{minipage} &
\begin{minipage}[t]{\linewidth}\vspace{0pt}\centering
\includegraphics{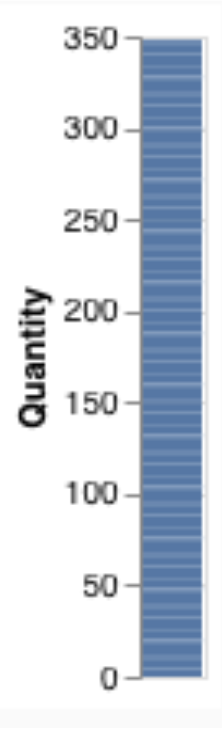}\vspace{2pt}\par
\RaggedRight\footnotesize A bar chart ordered by Quantity values for each Region; the description notes sorting by quantity but does not specify whether the order is ascending or descending.
\end{minipage} &
\begin{minipage}[t]{\linewidth}\vspace{0pt}\RaggedRight\footnotesize
Data Fidelity = \scorebadge{100\%}; Field Similarity = \scorebadge{100\%}; Chart Type Similarity = \scorebadge{100\%}; Axis Accuracy = \scorebadge{50\%}; Filter Accuracy = \scorebadge{100\%}; \textbf{Sort Accuracy = \scorebadge{0\%}}; Visual Encoding Accuracy = \scorebadge{100\%}.\\
\textbf{Overall Viz} \scorebadge{65\%}.\\
NL: Factual Grounding \scorebadge{60\%}; Assumptions Disclosure = \scorebadge{40\%}; Insightfulness = \scorebadge{20\%}; Coherence \scorebadge{80\%}.\\
\textbf{Overall NL} \scorebadge{50\%}.
\end{minipage} \\

2 &
\codewrap{\texttt{show me top accounts by attendees}} &
\begin{minipage}[t]{\linewidth}\vspace{0pt}\centering
\includegraphics{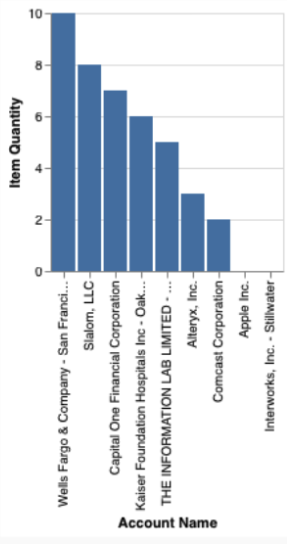}\vspace{2pt}\par
\RaggedRight\footnotesize A horizontal bar chart of the Top 10 Account Names, ranked by total (sum) Item Quantity and sorted from highest to lowest, making it easy to spot the highest-volume accounts.
\end{minipage} &
\begin{minipage}[t]{\linewidth}\vspace{0pt}\centering
\includegraphics[width=2in,keepaspectratio]{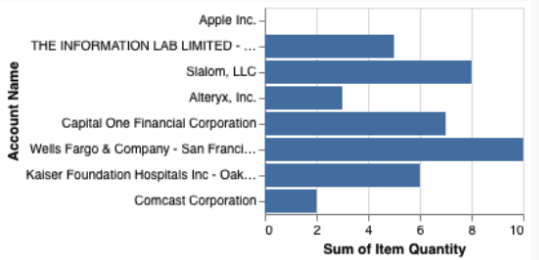}\vspace{2pt}\par
\RaggedRight\footnotesize A horizontal bar chart listing the Top 10 Account Names ranked by sum(Item Quantity), with bars sorted descending to highlight the largest contributors. (Accounts with null Item Quantity don’t contribute to the totals.)
\end{minipage} &
\begin{minipage}[t]{\linewidth}\vspace{0pt}\RaggedRight\footnotesize
Data Fidelity = \scorebadge{100\%}; Field Similarity \scorebadge{100\%}; Chart Type Similarity = \scorebadge{100\%}; Axis Accuracy = \scorebadge{100\%}; Filter Accuracy = \scorebadge{100\%}; Sort Accuracy = \scorebadge{100\%}; Visual Encoding Accuracy \scorebadge{95\%}.\\
\textbf{Overall Viz} \scorebadge{98\%}.\\
NL: Factual Grounding = \scorebadge{100\%}; Assumptions Disclosure = \scorebadge{100\%}; Insightfulness = \scorebadge{20\%}; Coherence \scorebadge{95\%}.\\
\textbf{Overall NL} \scorebadge{85\%}.
\end{minipage} \\

3 &
\codewrap{\texttt{count orders by categories}} &
\begin{minipage}[t]{\linewidth}\vspace{0pt}\centering
\includegraphics{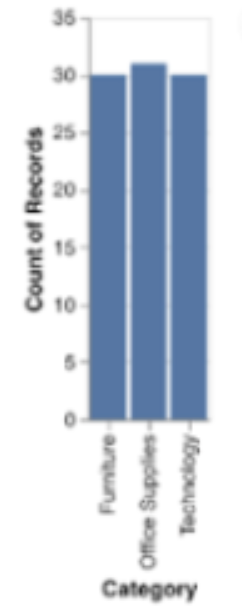}\vspace{2pt}\par
\RaggedRight\footnotesize A bar chart that counts orders per Category---the x-axis lists categories and the y-axis shows the count of Order IDs.
\end{minipage} &
\begin{minipage}[t]{\linewidth}\vspace{0pt}\centering
\includegraphics{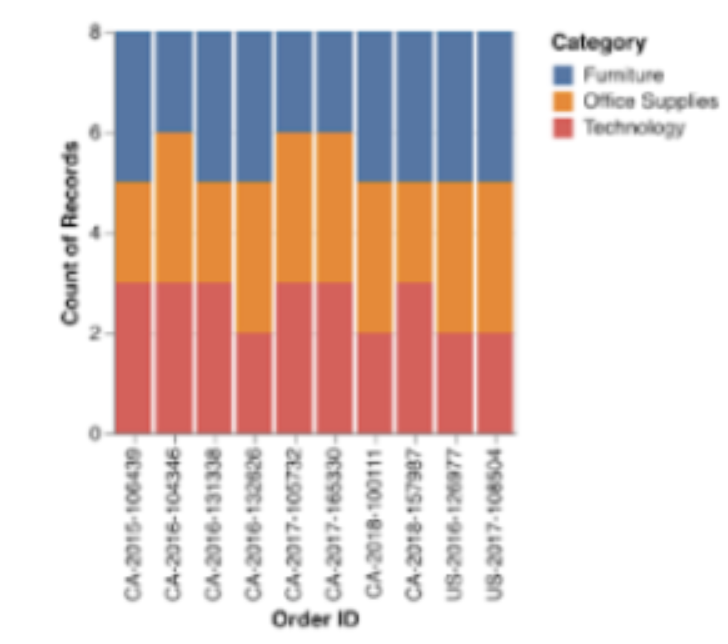}\vspace{2pt}\par
\RaggedRight\footnotesize A bar chart breaking down Order IDs by Category, with each category represented by a different color to distinguish groups.
\end{minipage} &
\begin{minipage}[t]{\linewidth}\vspace{0pt}\RaggedRight\footnotesize
Data Fidelity = \scorebadge{100\%}; Field Similarity = \scorebadge{100\%}; Chart Type Similarity \scorebadge{50\%}; Axis Accuracy = \scorebadge{100\%}; Filter Accuracy = \scorebadge{100\%}; Sort Accuracy = \scorebadge{100\%}; Visual Encoding Accuracy \scorebadge{85\%}.\\
\textbf{Overall Viz} \scorebadge{90\%}.\\
NL: Factual Grounding \scorebadge{60\%}; Assumptions Disclosure =\scorebadge{20\%}; Insightfulness = \scorebadge{20\%}; Coherence \scorebadge{90\%}.\\
\textbf{Overall NL} \scorebadge{50\%}.
\end{minipage} \\

4 &
\codewrap{\texttt{revenue versus earnings}} &
\begin{minipage}[t]{\linewidth}\vspace{0pt}\centering
\includegraphics[width=0.25in,keepaspectratio]{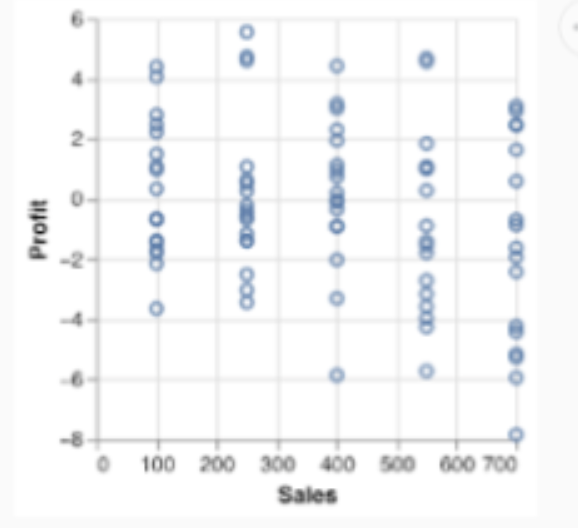}\vspace{2pt}\par
\RaggedRight\footnotesize A scatterplot with Sales on the x-axis and Profit on the y-axis, one point per record. Plotting raw values reveals how profit changes with sales.
\end{minipage} &
\begin{minipage}[t]{\linewidth}\vspace{0pt}\centering
\includegraphics[width=0.25in,keepaspectratio]{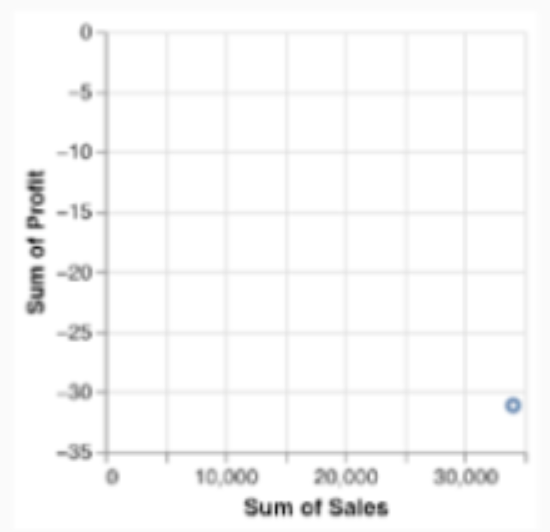}\vspace{2pt}\par
\RaggedRight\footnotesize A scatterplot comparing total Sales (x) to total Profit (y) so it collapses to a single point representing Sum of Sales vs. Sum of Profit.
\end{minipage} &
\begin{minipage}[t]{\linewidth}\vspace{0pt}\RaggedRight\footnotesize
Data Fidelity = \scorebadge{0\%}; Axis Accuracy = \scorebadge{95\%}; All other viz metrics = \scorebadge{100\%}.
\textbf{Overall Viz} =\scorebadge{85\%};\quad\\
NL: Factual Grounding \scorebadge{25\%}; Assumptions Disclosure = \scorebadge{0\%}; Insightfulness = \scorebadge{20\%}; Coherence \scorebadge{90\%}.\\
\textbf{Overall NL} \scorebadge{30\%}.
\end{minipage} \\
\bottomrule
\end{tabularx}
\caption{Worked examples showing how \lexara~metrics assess visualization and natural language quality responses interpretively.}
\label{tab:lexara-worked-examples}
\Description{This table shows four worked examples of user utterances, expected responses, model responses, and detailed metrics. Each row corresponds to a test case. In the first row, the utterance “Quantity on y-axis and Region on x-axis” is shown with expected and model bar charts of Quantity by Region. The expected response sorts regions by quantity, while the model output also shows a bar chart. Metrics such as Data Fidelity (100\%), Field Similarity (100\%), and Assumptions Disclosure (40\%) are listed. The second row shows “Sort by Quantity” with bar charts ordered by quantity; however, the model does not specify ascending or descending. Metrics show perfect axis accuracy but 0\% for Sort Accuracy. The third row shows “Show me top accounts by attendees” with expected and model horizontal bar charts listing top accounts, with some mismatches in ordering. The fourth row shows “Revenue versus earnings,” where the expected response is a scatterplot of Sales versus Profit per record, while the model produces an aggregated scatterplot collapsing all records into one point. Metrics reflect an aggregation mismatch with Data Fidelity at 0
Each row includes natural language descriptions of expected and actual charts, alongside metrics for visualization and language quality.}
\end{table*}
}

The practitioner evaluation criteria described in detail in \S\ref{sec:criteria} are implemented in code and their graded nature is illustrated through examples in this section. For examples of how \lexara~evaluates model responses using these metrics see Table \ref{tab:lexara-worked-examples}. The metrics are designed to operate over structured visualization grammars and natural language text, and do not assume vision inputs or explicit tool-calling support. To reflect the nuanced judgments practitioners make in evaluating CVA outputs, the metric scores are scaled to express partial correctness on a continuum rather than as binary outcomes. For instance, a model output that captures the general intent of the query but omits a key filter or misrepresents an encoding might receive a score of $70$ rather than $50$. This scaling is informed by formative user studies, where practitioners rated outputs on Likert-style scales and provided justifications reflecting degrees of acceptability. We anchor partial credit scores to thresholds that align with rubric-based assessment theory~\cite{Brookhart:2018} and graded benchmarking techniques~\cite{pathak:2025,hashemi-etal-2024-llm,atwood:2018} for supporting diagnostic debugging. This graded approach to benchmarking enables practitioners to distinguish between responses that are technically plausible but incomplete, versus those that are outright misleading or irrelevant.

\subsubsection{Visualization Quality Metrics}
Metrics systematically measure the quality of visualization responses and range from 0--100\%. Calculated by comparing the expected and actual visualization grammar spec and accommodating for partial credit or multiple plausible answers. 

\pheading{Data.} Refers to whether the visualization truthfully represents the underlying datasource. This includes checking for fidelity of rows, columns, and aggregations, as well as semantic alignment of selected fields with user intent.
    \begin{itemize}
        \item \textbf{Data Fidelity:} Checks if the visualization faithfully represents the underlying datasource (rows, columns, and processing like aggregations or means).  
            
            \begin{algorithmic}
            \State \textbf{function} \textsc{Score\_Data\_Fidelity}(expected, actual)
            \If{expected equals actual}
              \State \Return 100
            \ElsIf{same rows and fields(expected, actual) AND minor aggregation difference(expected, actual)}
              \State \Return 70
            \Else
              \State \Return 0
            \EndIf
            \State \textbf{end function}
            \end{algorithmic}

 \textit{Examples:} Expected: Sum of Quantity = 200 vs Actual: Count of Quantity = 200 $\rightarrow$ Score = \scorebadge{70\%} (aggregation mismatch). Expected: 100 rows vs Actual: 80 rows $\rightarrow$ Score = \scorebadge{0\%} (data missing).

        \item \textbf{Field Similarity:} Checks if bound fields match intended fields; partial credit if fields are semantically equivalent or have matching data types.
                    \begin{algorithmic}
                    \Function{score\_field\_similarity}{expectedSpec, actualSpec, datasourceMeta}
                      \State $E_x \gets$ expectedSpec.encoding.x.field
                      \State $E_y \gets$ expectedSpec.encoding.y.field
                      \State $A_x \gets$ actualSpec.encoding.x.field
                      \State $A_y \gets$ actualSpec.encoding.y.field
                    
                      \Function{canon}{f, meta} 
                        \State names $\gets \{\, meta[f].name \,\} \cup$ meta[f].aliases
                        \State \Return stem(lowercase(join(names)))
                      \EndFunction
                    
                      \Function{cosSimStems}{s1, s2}
                        \State $\vec{v}_1 \gets$ bow(s1); \quad $\vec{v}_2 \gets$ bow(s2)
                        \If{$\|\vec{v}_1\| = 0$ or $\|\vec{v}_2\| = 0$} 
                          \State \Return $0$
                        \EndIf
                        \State \Return $\dfrac{\vec{v}_1 \cdot \vec{v}_2}{\|\vec{v}_1\| \|\vec{v}_2\|}$
                      \EndFunction
                    
                      \State $S_x \gets$ cosSimStems(canon($E_x$, datasourceMeta), canon($A_x$, datasourceMeta))
                      \State $S_y \gets$ cosSimStems(canon($E_y$, datasourceMeta), canon($A_y$, datasourceMeta))
                      \State $S \gets (S_x + S_y)/2$
                    
                      \State $T_x \gets$ [meta[$E_x$].dataType = meta[$A_x$].dataType]
                      \State $T_y \gets$ [meta[$E_y$].dataType = meta[$A_y$].dataType]
                      \State bonus $\gets 10$ if $(T_x \wedge T_y)$ else $0$
                    
                      \State \Return $\min(100,\; 100 \times S + bonus)$
                    \EndFunction

                 \end{algorithmic}
            \textit{Examples:} Related data type and semantic fields: \texttt{Order Date vs Ship Date} $\rightarrow$ base semantic similarity $0.77$; both temporal $\Rightarrow$ + 10\% type bonus. Score: $\min(100\%,\;77\% + 10\%) =$ \scorebadge{87\%}
            
              Unrelated data types and semantic fields: \texttt{Region vs Category} $\rightarrow$ base semantic similarity $0.29$; both discrete dimensions $\Rightarrow$ +10\% type bonus. Score: $\min(100\%,\; 29\% + 10\%) =$ \scorebadge{39\%}
            
              Missing axis: \texttt{Sales vs (missing Y axis)} $\rightarrow$ no comparable field, similarity $=0.0$; no type bonus. Score: \scorebadge{0\%}
            
            \noindent\emph{Note:} For two-axis charts, compute $S_x$ and $S_y$ separately via stemmed cosine similarity, set $S=(S_x+S_y)/2$, then add a single 10\% bonus only if both axis data types match; cap at $100\%$.
    \end{itemize}
    
\pheading{Chart Type.} measures how well the model’s chosen mark type aligns with the mark types that Tableau’s Show Me engine \cite{showme} recommends for the same set of data fields. We (i) run Show Me on the expected fields to obtain a ranked list of recommended chart/mark types, then (ii) compare the model’s chosen mark against that list:
            \begin{algorithmic}
            \Function{score\_chart\_similarity}{expectedSpec, actualSpec, datasourceMeta}
              \State $F \gets \{\, expectedSpec.encoding.x.field, \; expectedSpec.encoding.y.field \,\}$
              \State $R \gets$ ShowMeRecommend($F$, datasourceMeta) \Comment{ranked list of mark types}
              \State $M \gets$ normalizeMark(actualSpec.mark) \Comment{e.g., ``line'', ``bar'', ``area'', ``point'', ``table''}
            
              \If{$R = \emptyset$}
                \State \Return $0\%$
              \ElsIf{$M = R[1]$} \Comment{top recommendation}
                \State \Return $100\%$
              \ElsIf{$M \in R$} \Comment{other Show Me recommendation}
                \State \Return $50$
              \Else
                \State \Return $0\%$
              \EndIf
            \EndFunction
            \end{algorithmic}

 \textit{Examples:} Best recommended chart: Data = time series, Model = \texttt{line} $\rightarrow$ Score = \scorebadge{100\%}. Plausible but not best: 
                    Data = time series, Model = \texttt{area} $\rightarrow$ Score = \scorebadge{50\%}. Alternate plausible: 
                    Data = category + measure, Model = \texttt{pie} $\rightarrow$ Score = \scorebadge{50\%}. Poor choice:
                    Data = two measures, Model = \texttt{table} $\rightarrow$ Score = \scorebadge{0\%}. 

\pheading{Functionality} Captures whether the visualization operates correctly in terms of axes, filters, and sorts. A functionally correct chart reflects accurate axis assignments, scale and baselines, appropriate filtering, and expected ordering. 
    \begin{itemize}
        \item \textbf{Axis Accuracy.}
Evaluates whether the X/Y axes use the \emph{intended fields} and \emph{parameters}, giving graded credit for semantic proximity of field names and incorporating data-type compatibility directly into the similarity score.

\begin{algorithmic}
\Function{score\_axis\_accuracy}{expectedSpec, actualSpec, datasourceMeta}
  \State $E_x \gets$ expectedSpec.encoding.x.field;\;\; $E_y \gets$ expectedSpec.encoding.y.field
  \State $A_x \gets$ actualSpec.encoding.x.field;\;\;\;\;\;\;\, $A_y \gets$ actualSpec.encoding.y.field

  \State $S_x \gets$ cosSimStems(canon($E_x$, datasourceMeta), canon($A_x$, datasourceMeta))
  \State $S_y \gets$ cosSimStems(canon($E_y$, datasourceMeta), canon($A_y$, datasourceMeta))

  \State $T_x \gets \mathbf{1}_{(meta[E_x].dataType = meta[A_x].dataType)}$
  \State $T_y \gets \mathbf{1}_{(meta[E_y].dataType = meta[A_y].dataType)}$

  \State $S_x' \gets 0.9 \times S_x + 0.1 \times T_x$
  \State $S_y' \gets 0.9 \times S_y + 0.1 \times T_y$

  \State $S \gets (S_x' + S_y')/2$
  \State $score \gets 100 \times S$

  \If{axesSwapped(expectedSpec, actualSpec)}
    \State $score \gets 0.5 \times score$
  \EndIf

  \If{wrongScaleOrBaseline(actualSpec)}
    \State $score \gets 0.7 \times score$
  \EndIf

  \State \Return $\min(100,\; score)$
\EndFunction
\end{algorithmic}

        \item \textbf{Filter Accuracy}
            Measures how well applied filters match the expected set, allowing partial credit when \emph{field names} are semantically close and \emph{values} normalize to the same concept; adds a small bonus when matched fields share data types. \textit{Examples:} Extra filter: expected \{Year=2023\}, actual \{Year=2023, Region=West\} $\rightarrow$ \scorebadge{50\%}  (one match over soft union), $+10\%$ if types align; Semantic field/value match: expected \{Region=West\}, actual \{SalesRegion=West\} $\rightarrow$ high semantic similarity and added value equivalence $\Rightarrow $ \scorebadge{100\%}; Wrong value: expected \{Month=Jan\}, actual \{Month=Feb\} $\rightarrow$ \scorebadge{0\%}
            
            \begin{algorithmic}
\Function{score\_filter\_accuracy}{expectedSpec, actualSpec, meta}

  \State $E \gets$ normalizeFilters(expectedSpec.transform.filter)
  \State $A \gets$ normalizeFilters(actualSpec.transform.filter)

  \State matched $\gets 0$
  \State usedA $\gets \emptyset$
  \State usedE $\gets \emptyset$

  \For{each $e \in E$}
    \State best $\gets 0$
    \State bestIdx $\gets$ none

    \For{each $a \in A$ not in usedA}

      \State $sf \gets$
      \State \hspace{1.5em}
        cosSimStems(
      \State \hspace{3em}
        canon($e.field$, meta),
      \State \hspace{3em}
        canon($a.field$, meta))

      \If{$sf \ge \tau_f$ \textbf{and} valuesEquivalent($e$, $a$)}
        \State $sim \gets (sf + opMatch($e$, $a$))/2$
        \If{$sim > best$}
          \State best $\gets sim$
          \State bestIdx $\gets a$
        \EndIf
      \EndIf

    \EndFor

    \If{bestIdx $\neq$ none}
      \State matched $\gets matched + best$
      \State usedA $\gets usedA \cup \{bestIdx\}$
      \State usedE $\gets usedE \cup \{e\}$
    \EndIf

  \EndFor

  \State matchCount $\gets |usedE|$
  \State unionSize $\gets |E| + |A| - matchCount$

  \If{unionSize = 0}
    \State \Return $100$
  \EndIf

  \State base $\gets 100 \times
        \dfrac{matched}{unionSize}$

  \State base $\gets \min(100,\, base)$

  \State typesAgree $\gets$
  \State \hspace{1.5em} allMatchedTypesEqual(usedE, usedA)

  \If{typesAgree}
    \State bonus $\gets 10$
  \Else
    \State bonus $\gets 0$
  \EndIf

  \State \Return $\min(100,\, base + bonus)$

\EndFunction
\end{algorithmic}

        \item \textbf{Sort Accuracy:} Assesses whether sort \emph{fields} and \emph{directions} match, granting graded credit for semantic proximity of the sort key and a type-consistency bonus.

            \begin{algorithmic}
            \Function{score\_sort\_accuracy}{expectedSpec, actualSpec, datasourceMeta}
              \State $E \gets$ expectedSpec.sort \Comment{(field, direction)}
              \State $A \gets$ actualSpec.sort \Comment{(field, direction)}
            
              \If{$E$ is none \textbf{ and } $A$ is none} 
                \State \Return $100$
              \EndIf
              \If{$E$ is none \textbf{ xor } $A$ is none} 
                \State \Return $0$
              \EndIf
            
              \State $S_f \gets$ cosSimStems(canon($E.field$, meta), canon($A.field$, meta))
            
              \State $D \gets$
                \begin{tabular}[t]{@{}l@{}}
                  $1.0$ if $E.dir = A.dir$;\\
                  $0.5$ if $E.dir \neq A.dir$;\\
                  $0$ if $A.dir$ missing when $E.dir$ specified
                \end{tabular}
            
              \State $base \gets 100 \times S_f \times D$
              \State $bonus \gets 10$ if meta[$E.field$].dataType $=$ meta[$A.field$].dataType else $0$
            
              \State \Return $\min(100,\; base + bonus)$
            \EndFunction
            \end{algorithmic}

            \textit{Examples:} Right field, wrong direction: Sales  \texttt{desc}  vs. Sales  \texttt{asc}  $\rightarrow$ $S_f = 1.0$, $D=0.5$; $50\%$ (+$10\%$ if type matches) $\Rightarrow$ up to \scorebadge{60\%}; Semantic key: Revenue  \texttt{desc}  vs. SalesAmount  \texttt{desc}  (treated as same concept) $\rightarrow$ $S_f = 0.8$, $D=1.0$, type bonus $=10$ $\Rightarrow =$ \scorebadge{100\%}; Missing sort: expected sort  \texttt{desc} , actual none $\rightarrow$ \scorebadge{0\%}

    \end{itemize}

\pheading{Design.} Focuses on how clearly and meaningfully information is encoded. This involves assessing the accuracy of visual encodings (e.g., color, size, labels), the appropriateness of design choices for interpretability, and whether interactive elements (e.g., tooltips) provide correct contextual information.
    \begin{itemize}
\item \textbf{Visual Encoding Accuracy.}
Measures how truthfully and clearly the chart maps data to visual channels,
using a graded score per channel (\emph{color, shape, opacity, text, size}),
then averaging across channels. For each channel we combine:
(i) semantic match of bound field,
(ii) data-type consistency, and
(iii) design best-practice adherence
(e.g., hue for nominal, gradient for quantitative, contrast, legibility).

\begin{algorithmic}
\Function{score\_encoding\_accuracy}{expectedSpec, actualSpec, meta}

  \State channels $\gets \{color, shape, opacity, text, size\}$
  \State scores $\gets [\,]$

  \For{each $c$ in channels}

    \State $E \gets$ expectedSpec.encoding.$c$
    \State $A \gets$ actualSpec.encoding.$c$

    \If{$E$ is none \textbf{and} $A$ is none}
      \State append $100$ to scores
      \State \textbf{continue}
    \EndIf

    \State presence $\gets 0$
    \If{$E$ and $A$ present}
      \State presence $\gets 100$
    \ElsIf{$A$ present}
      \State presence $\gets 50$
    \EndIf

    \State sem $\gets 0$
    \If{$E$ and $A$ present}
      \State sem $\gets 100 \times$
      \State \hspace{1.5em}
        cosSimStems(
      \State \hspace{3em}
        canon($E.field$, meta),
      \State \hspace{3em}
        canon($A.field$, meta))
    \EndIf

    \State typeOK $\gets 0$
    \If{$E$ and $A$ present}
      \If{meta[$E.field$].dataType =
          meta[$A.field$].dataType}
        \State typeOK $\gets 100$
      \EndIf
    \EndIf

    \State practice $\gets 100 \times$
    \State \hspace{1.5em}
      bestPracticeScore($c$, $E$, $A$, meta)

    \State $s_c \gets$
    \State \hspace{1.5em}
      $0.3 \cdot presence +$
    \State \hspace{1.5em}
      $0.4 \cdot sem +$
    \State \hspace{1.5em}
      $0.1 \cdot typeOK +$
    \State \hspace{1.5em}
     $ 0.2 \cdot practice$

    \State append $s_c$ to scores

  \EndFor

  \State \Return mean(scores)

\EndFunction
\end{algorithmic}

            \textit{Examples:}
Exact, well-designed: Color = Region (nominal) with categorical palette, text labels on bars, size not used $\rightarrow$ Score \scorebadge{100\%}; Good semantics, minor design gap: Color = Sales (quantitative) but uses categorical palette (should be gradient) $\rightarrow$ Score \scorebadge{80\%}; Alternative acceptable: Expected no opacity, actual uses opacity to mitigate overplotting in dense scatter (quantitative) 
                        $\rightarrow$ Score \scorebadge{80\%}; 
Mismatched channel: Size = Region (nominal) with many categories 
                        $\rightarrow$ Score \scorebadge{40\%}; 
Missing expected channel: Expected color by Region, actual has no color encoding 
                        $\rightarrow$ Score = \scorebadge{0\%} for color (averaged across channels)

        \item \textbf{Interactivity Accuracy}
            Scores how well interactive affordances (e.g., tooltips, selection, zoom/pan, drill-down) support accurate, relevant, and usable reading. 
            Allows \emph{multiple correct answers}: if the actual design includes an alternate but reasonable set of fields or interactions that covers the required information, it receives partial credit. We combine: (i) \textit{coverage} of required info, (ii) \textit{correctness} of shown values/aggregations, and (iii) \textit{usability} best practices (formatting, units, concise lists, interaction consistency).
            \begin{algorithmic}
\Function{score\_interactivity\_accuracy}{expectedSpec, actualSpec, datasourceMeta}
  \State $R \gets$ requiredTooltipFields(expectedSpec) 
    \Comment{fields bound to x,y,color,size + key filters/measures}
  \State $A \gets$ actualTooltipFields(actualSpec)
  \State $M \gets$ matchFieldsSemantically($R$, $A$, datasourceMeta) 
    \Comment{one-to-one best semantic matches}

  \State coverage $\gets 100 \times \dfrac{|M|}{\max(1, |R|)}$ 
    \Comment{fraction of required info covered}
  \State correctness $\gets 100 \times$ mean(\textsc{aggOK}($m$) for $m \in M$) 
    \Comment{aggregations/values align with spec}
  \State extras $\gets 100 \times$ normalized count of additional relevant fields in $A \setminus M$
  \State redundancyPenalty $\gets 100 \times$ penalty for duplicates/noise (e.g., repeating same measure twice)

  \State tooltipScore $\gets$ clamp\Big( 
    $0.6 \cdot$coverage $+ 0.3 \cdot$correctness $+ 0.1 \cdot$extras $- 0.1 \cdot$redundancyPenalty \Big)

  \State interactionsExpected $\gets$ interactionsFrom(expectedSpec) 
    \Comment{e.g., selection, zoom, drill}
  \State interactionsActual $\gets$ interactionsFrom(actualSpec)

  \State interMatch $\gets 100 \times$ softJaccard(interactionsExpected, interactionsActual) 
    \Comment{partial credit for alternates}
  \State consistency $\gets 100 \times$ \textsc{interactionConsistencyOK}()
    \Comment{tooltips/selection respect filters/sorts/encodings}
  \State usability $\gets 100 \times$ avg(\textsc{formattingOK}(), \textsc{unitsShown}(), \textsc{brevityOK}()) 
    \Comment{avoid long lists, show units}

  \State $w_t \gets 0.6,\; w_i \gets 0.2,\; w_c \gets 0.1,\; w_u \gets 0.1$
  \State \Return $w_t \cdot$tooltipScore $+ w_i \cdot$interMatch $+ w_c \cdot$consistency $+ w_u \cdot$usability
\EndFunction
\end{algorithmic}
            \noindent \textit{Examples.}

Complete \& correct tooltips: Show X, Y, color field, and measure with correct aggregation, formatted with units; selection highlights series consistently $\rightarrow$ Score \scorebadge{100\%}. Alternate but acceptable: Expected \{Month, Sales, Region\}, actual shows \{Month, SalesAmount, RegionName, Profit\}; semantic matches for required fields plus a relevant extra; formatting OK $\rightarrow$ Score \scorebadge{95\%}. Partially correct:  Shows X/Y but omits filter context and uses count instead of sum in tooltip $\rightarrow$ Score \scorebadge{60\%}. Redundant/noisy: Long tooltip lists with duplicate fields, inconsistent units 
$\rightarrow$ Score \scorebadge{50\%}. Missing: No tooltips or interactive affordances when expected $\rightarrow$ Score = \scorebadge{0\%}.

\end{itemize}

\subsubsection{Natural Language Response Quality}
These metrics evaluate natural language responses in CVA. Four out of five of them are implemented as LLM-as-a-Judge metrics grounded in human-in-the-loop evaluations from Formative Study 2 [§\ref{sec:formative}.2], where participants rated model outputs turn-by-turn and articulated why certain responses were accurate, incomplete, or misleading. These qualitative judgments provided a rich, annotated dataset used to craft few-shot rubrics for automated scoring. For example, vague or contradictory explanations were labeled by participants as ``low coherence," while detailed responses naming filters, time frames, and assumptions were judged ``high on assumptions disclosure.'' By embedding these annotated examples into the judge prompts, we ensured that LLM-based ratings would better align with practitioner expectations, capture graded correctness, and remain interpretable. This approach also gave end-users a traceable influence on the design of automated evaluation, supporting transparency and trust in the resulting metrics.
    
\pheading{Factual Grounding:} This is calculated programmatically and ensures the explanation conveys the semantically similar facts as the visualization (measures, magnitudes, directions). 
        \begin{algorithmic}
\Function{score\_factual\_grounding}{expectedText, actualText}

  \State $e \gets$ embedding(expectedText)
  \State $a \gets$ embedding(actualText)

  \State similarity $\gets$
  \State \hspace{1.5em} cosine($e$, $a$)

  \If{contradiction(expectedText, actualText)}
    \State \Return $0$
  \Else
    \State \Return $100 \times similarity$
  \EndIf

\EndFunction
\end{algorithmic}

\textit{Examples:}
            \begin{itemize}
                \item Expected: ``Profit climbed 8\% year-over-year'' vs Actual: ``Profit up eight percent year-over-year'' $\rightarrow$ Score = \scorebadge{100\%}.
                \item Expected: ``Profit climbed 8\% year-over-year'' vs Actual: ``Profit improved year-over-year'' $\rightarrow$ Score = \scorebadge{70\%} (magnitude missing).
                \item Expected: ``Profit climbed 8\% year-over-year'' vs Actual: ``Revenue grew 8\%'' $\rightarrow$ Score = \scorebadge{0\%} (wrong measure).
            \end{itemize}

\pheading{Analytical Thinking:} 
        \begin{itemize}
            \item \textbf{Assumptions Disclosure: \footnote{This is an LLM-as-a-Judge Metric, please see the human annotated few-shot examples used to prompt this in the supplementary materials.}} Evaluates whether the response surfaces relevant assumptions (filters, time frames, aggregation choices).    
             \textit{Examples:}
            \begin{itemize}
            \item Actual: ``This assumes the Region filter is set to North America and values are aggregated monthly'' $\rightarrow$ Score = 4 (relevant assumptions).
            \item Actual: ``These insights assume data excludes returns, is filtered to 2023, and that Sales reflects total revenue, not net'' $\rightarrow$ Score = 5 (comprehensive).
            \end{itemize}
            \item \textbf{Insightfulness:}\footnotemark[1] Captures depth of analysis, identifying trends, exceptions, and actionable implications.  
             \textit{Examples:}
             \begin{itemize}
                \item Actual: ``Sales increased'' $\rightarrow$ Score =2 (basic observation).
                 \item Actual: ``From Q1 to Q4, Electronics in the West grew 25\%, Apparel in the South fell 10\%, suggesting a shift in seasonal demand'' $\rightarrow$ Score = 5 (rich, actionable).
                 \end{itemize}
        \end{itemize}

\pheading{Conversation Quality: }
        \begin{itemize}
            \item \textbf{Coherence:\footnotemark[1]} Evaluates whether the response is internally consistent and logically structured.  
             \textit{Examples:}

 Actual: ``Sales are up, but that means profit is lower, so we should cut inventory'' $\rightarrow$ Score = 1 (contradictory).
Actual: ``Inventory is down. Sales are good. Profit is low. Maybe a trend?'' $\rightarrow$ Score = 2 (disorganized).
Actual: ``Sales increased, possibly leading to higher profit. Inventory dropped, which might be a concern'' $\rightarrow$ Score = 3 (mostly coherent).
Actual: ``Sales rose in Q4, contributing to higher profits. Inventory dropped significantly, which could create supply issues next quarter'' $\rightarrow$ Score = 4 (well-structured).
Actual: ``Q4 sales increased 20\%, profits rose 15\%, while inventory declined 30\%, raising fulfillment concerns for Q1'' $\rightarrow$ Score = 5 (clear and precise).

            \item \textbf{Follow-up Relevance:\footnotemark[1]} Checks whether the response remains grounded in prior turns of the conversation (multi-turn only).  
             Examples: 

                \item \begin{quote}
                \textsc{Previous user utterance:} ``Focus on high-growth segments.'' \\
                \textsc{Response:} ``I included segment data like you asked earlier.'' \\
                $\rightarrow$ Score = 2 (minimal linkage).
                \end{quote}
     
                \item \begin{quote}
                \textsc{Previous user utterance:} ``Focus on high-growth segments in Q3 only.'' \\
                \textsc{Response:} ``This line chart filters to Q3 only, shows Technology outperforming all others, continuing the trend we saw last week.'' \\
                $\rightarrow$ Score = 5 (fully grounded).
                \end{quote}
            \end{itemize}

\subsection{\lexara's Interactive CVA Evaluation Tool}
\label{sec:tool}
\begin{figure*}
\includegraphics[width=\textwidth]{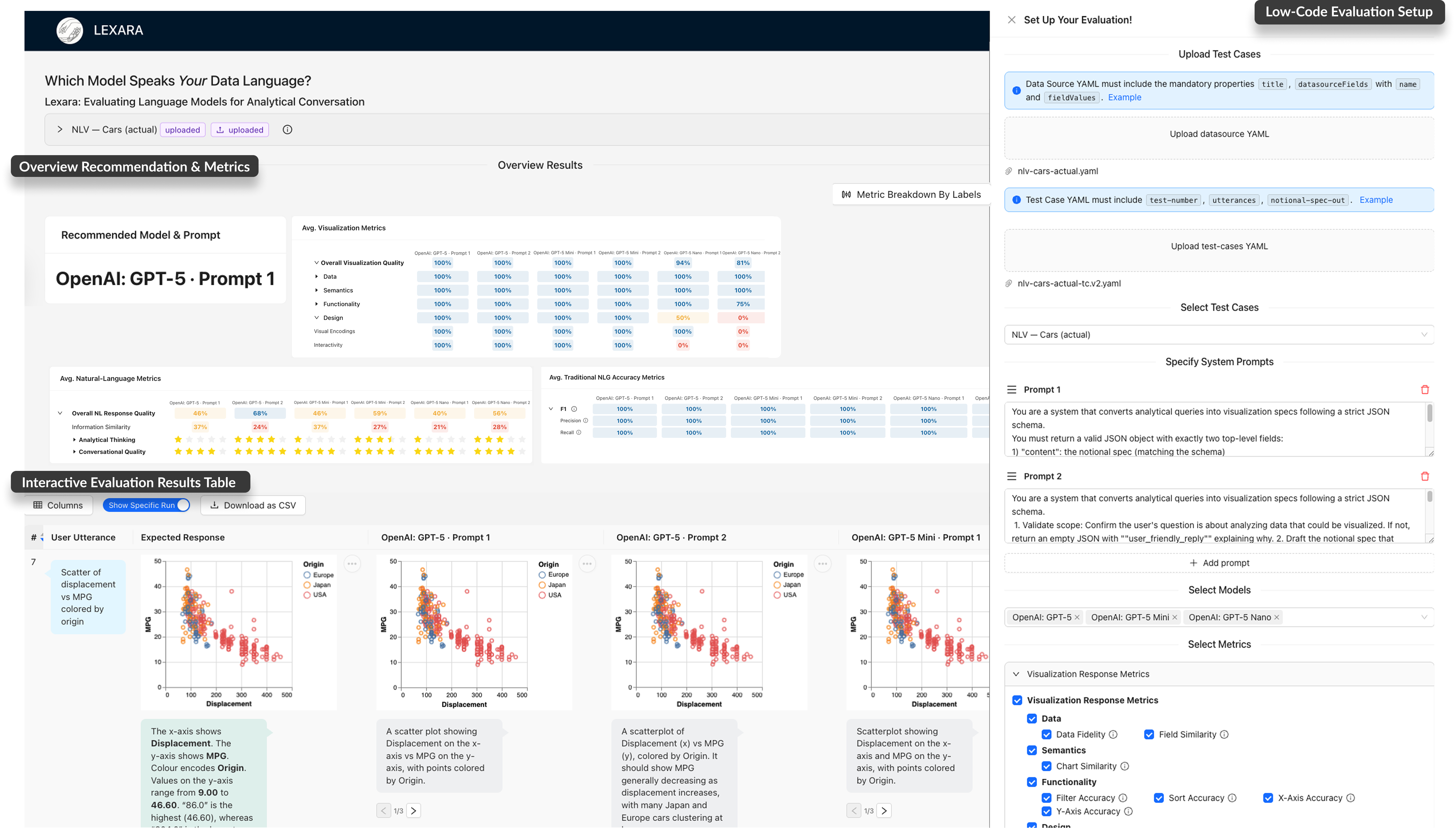}
\caption{\lexara's interactive CVA evaluation interface supports two core workflows: (1) an \textbf{Evaluation Setup Panel} where practitioners upload datasources, define test cases, specify prompts, models, expected outputs, and configure CVA-specific metrics; and (2) an \textbf{Interactive Results Table} that streams model outputs—visualizations, structured specs, and natural language—side-by-side. The table enables multi-granular inspection, with expandable metric categories, on-hover explanations, and tools to trace divergences between expected and actual outputs.} 
\label{fig:setupEval}
\vspace{-5mm}
\end{figure*} 
\Description{This screenshot shows the Lexara interface for interactive evaluation. On the far right, the “Evaluation Setup Panel” allows evaluators to upload test cases, datasources, and prompts, and to select models and metrics. In the center, the ``Evaluation Results Table'' displays side-by-side outputs from different models and prompts for the same utterance. For example, the user request “Scatter of displacement vs MPG colored by origin” is shown with the expected chart and outputs from GPT-5 variants, each rendered as scatterplots with color encodings. At the top of the page is the “Overview Recommendation and Metrics” panel, which lists overall visualization and natural language metrics for each model–prompt combination. Key sections are labeled with callouts: the top panel highlights recommendations and aggregate metrics, while the bottom interactive table supports inspecting outputs side-by-side. This figure illustrates how evaluators configure experiments and view results in real time.}

Building on the real-world test cases [\S\ref{sec:testcases}] and user-centered metrics [\S\ref{sec:metrics}], effective CVA evaluation also requires an interactive tool that simplifies setup, supports low-/no-code use, and surfaces actionable insights. The \lexara~interface is designed to address fragmented workflows, opaque outputs, and the disconnect between aggregate metrics and specific failures, supporting our design considerations [D1–D7]. It enables users to run benchmarks on custom data and prompts, compare multi-format outputs side by side, and drill down from high-level metrics to turn-level diagnostics.

\subsubsection{Evaluation Setup} (Figure \ref{fig:setupEval})
\lexara~enables low-code practitioners to run systematic CVA benchmarking experiments across multiple models and system prompts \textbf{[D1–D3]}. The setup workflow includes intentional defaults, helpful templates, and guardrails for error handling. Each run treats a configuration of one or more models and system prompts applied to a shared set of test cases as the object of evaluation. That is, the toolkit focuses on evaluating model–prompt behavior in the CVA backend (e.g., datasource interpretation, visualization specification, explanation quality), rather than instrumenting every component of a deployed CVA application such as the UI, logging pipeline, or enterprise orchestration.


\pheading{Upload datasource and test case files.}  Practitioners can upload their own datasources and test case files, or select from a sample in the 'Select Test Case' dropdown. To create their own files, they are given guidance on template with required fields, structure, and an example file stub. When the uploaded files do not match requirements, the error messages are specific in explaining how to fix the issue. Upon successful upload, they can preview the datasource table and test cases in the Evaluate Test Cases table where each test case ID is a conversation, each row is a user utterance in the conversation with labels, clicking on `+` next to the test case ID unfurls all the rows/utterances in multi-turn conversations. 

\pheading{Specify system prompts.} To compare prompt variants, practitioners can define multiple system prompts using an example prompt template with required variables (e.g., \texttt{datasource}, \texttt{utterance}, and expected \texttt{JSON visualization grammar structure}).  Prompts are numbered for traceability in results. 

\pheading{Select Models.} The toolkit currently supports 10 models: the latest OpenAI (\texttt{GPT-5, GPT-5-mini, GPT-5-nano, o3, o4-mini}), Anthropic (\texttt{claude-opus-4, claude-3.7-sonnet}, and Deepseek models (\texttt{r1}). These were selected as commonly used models by participants in the formative studies. Practitioners are encouraged to bring their own API keys for models. 
In our current implementation, \lexara~treats all evaluated models as text-only chat endpoints: we send natural language utterances and receive JSON visualization specifications plus textual explanations. Extending the toolkit to exercise models’ full multimodal and tool-use capabilities is left as future work (see §7.2).

\pheading{Select Metrics.} \textbf{[D7]}  Practitioners can select specific visualization and natural response quality metrics, or traditional natural language metrics (F1, Precision, Recall). Each metric has a tooltip with the definition (see \S\ref{sec:metrics} for details). \textbf{LLM-as-a-Judge Recommendation:} For metrics that require LLM-as-a-Judge, \lexara~recommends one by annotating the model in the drop-down with \texttt{(recommended)} next to its name. This recommendation is made heuristically by following best practices to reduce bias \cite{Zheng2023MTBench}: selecting the strongest model outside the LLM families of models getting evaluated to reduce self-bias. The judge models are instructed to ignore style or truncate or equalize answer length to avoid verbosity bias. Furthermore, to align the Judge with practitioners' evaluation criteria, we apply few-shot learning by sharing examples of ratings distilled from the formative study with end-users. 

\pheading{Specify Test Cases.} Practitioners can specify individual test cases or contiguous ranges of test case IDs. Leaving this blank will execute all test cases in the file. 

\pheading{Specify Number of Runs.} By default, \lexara~executes three replications per (model $\times$ system-prompt $\times$ judge) configuration to reduce run-to-run variance. Practitioners may adjust the number of replications to 1–5: set $1$ for exploratory spot checks and up to $5$ for increased reliability during benchmarking. Practitioners can examine the results from each run as they stream in to the Evaluation Results table. 

Pressing the \textit{Evaluate} button initiates an evaluation experiment, while the Stop button immediately terminates the ongoing evaluation.

\subsubsection{Interactive Test Cases Table} \textbf{[D4-6]}  As evaluations run, the table dynamically populates with multi-format responses, visualizations, natural language outputs, and JSON specs for each model $\times$ prompt combination. Corresponding metrics for visualization quality, language accuracy, and traditional Natural Language Generation (NLG) scores are computed in real time. To support focused analysis, the table offers spreadsheet-like features: columns can be filtered, hidden, or frozen, enabling flexible, side-by-side comparisons.

\begin{figure*}[!t]
\includegraphics[width=\textwidth]{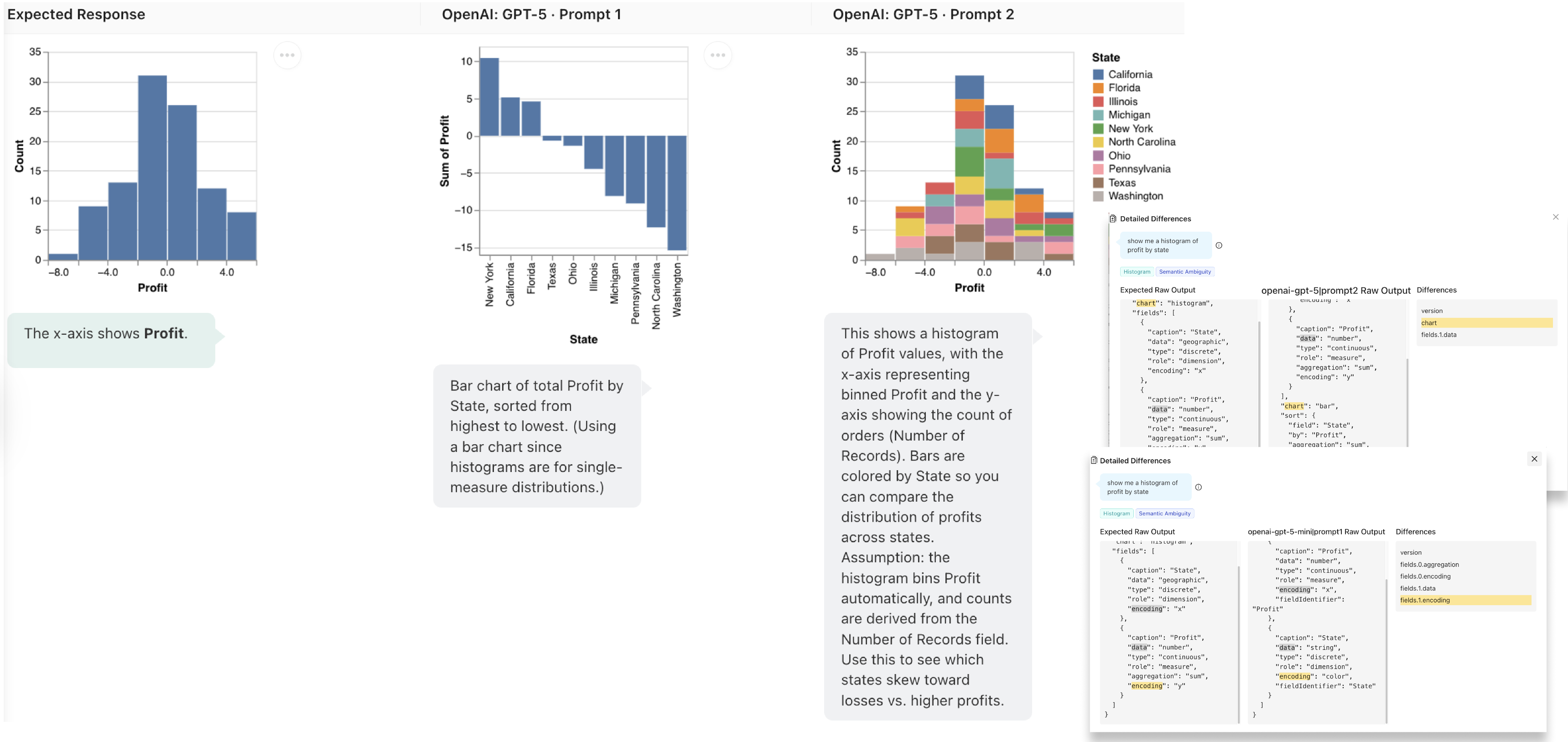}
\caption{For each user request, the system aligns expected and actual outputs across three formats: visualizations, natural language explanations, and JSON specifications. By surfacing detailed differences (e.g., encodings, aggregations, chart types), the interface enables practitioners pinpoint divergences, understand model behavior, and diagnose strengths or failure modes for various analytic tasks.} 
\label{fig:multiformat-outputs}
\vspace{-5mm}
\end{figure*}

\textbf{Multi-format Response Cells} (Figure \ref{fig:multiformat-outputs}) display each model's response as an interactive Vega-Lite chart and accompanying natural language explanation. These are rendered using a custom engine that transforms high-level JSON specs into Vega-Lite, allowing direct visual comparison with expected outputs.

\Description{This figure compares expected and model-generated outputs for a given user utterance. On the left, the expected response is a simple histogram of Profit, with the x-axis showing binned profit values and the y-axis showing counts. In the middle, GPT-5 Prompt 1 produces a bar chart of total Profit by State, sorted highest to lowest, while Prompt 2 produces a multi-colored histogram of Profit by State, with each bar colored by a different state. On the right, detailed JSON specifications are shown alongside difference panels. These highlight mismatches between the expected and actual grammar specifications, such as aggregation types or encodings. Below the visualizations, natural language explanations describe the outputs, pointing out assumptions and differences. This figure demonstrates how Lexara enables evaluators to align visual, textual, and specification-level outputs and inspect divergences.}

\begin{figure*}[!t]
\includegraphics[width=\textwidth]{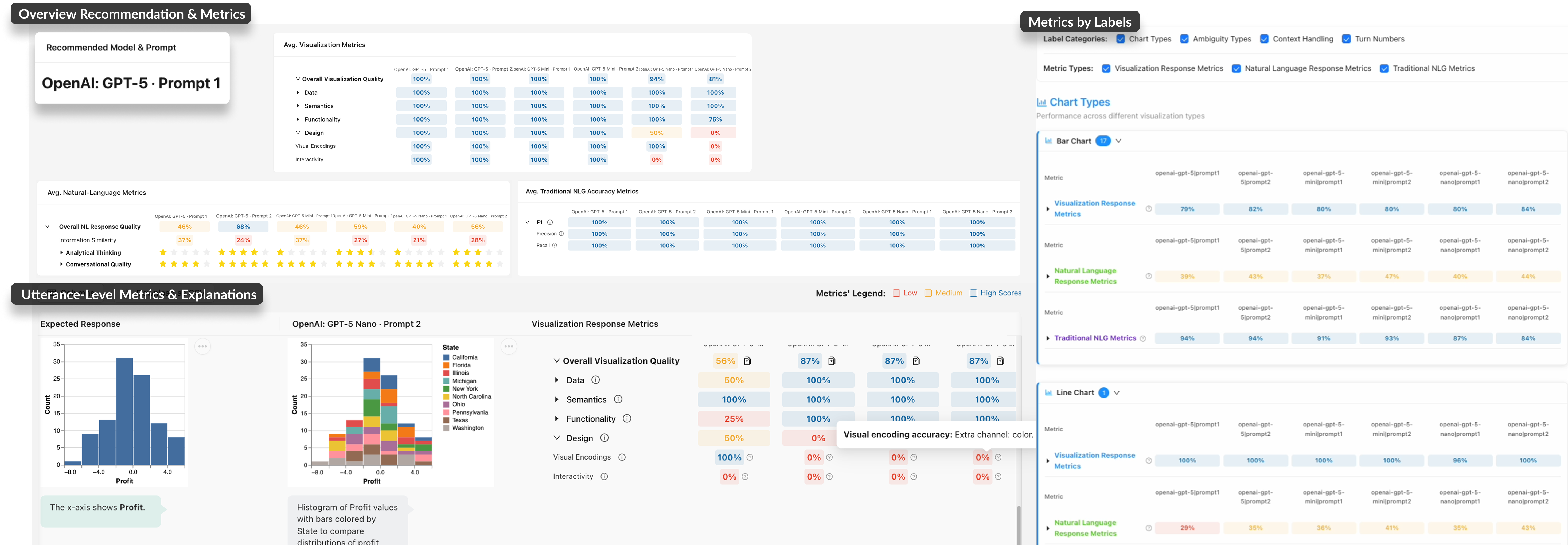}
\caption{The overview panel (top left) highlights recommended model–prompt pairs and aggregated metrics. The label view (top right) breaks down results by chart type, ambiguity, and context-handling. The utterance-level view (bottom) contrasts expected vs. actual responses with detailed metric explanations.} 
\label{fig:multilevel-metrics}
\vspace{-5mm}
\end{figure*}

\textbf{Hierarchical Metrics Drill-Down Cells} (Figure \ref{fig:multilevel-metrics}) Each column represents a metric category: Visualization Response, Natural Language Response, and Traditional NLG. A color-coded scale (red = low, yellow = mid, blue = high) highlights performance at a glance. The drill-down follows an overview + detail-on-demand pattern. For example, the Visualization column initially shows an overall quality score. Expanding reveals subcategories like \textit{Data}, \textit{Semantics}, \textit{Functionality}, and \textit{Design}, which further break down into granular metrics (e.g., \textit{Data Fidelity}, \textit{Sort Accuracy}, \textit{Visual Encodings}). This layered design supports rapid scanning with selective deep dives. Interactive explanations enhance interpretability: hovering on a score reveals the expected vs. actual output (e.g., \textit{Sort: Expected descending}, \textit{Model: none}). For LLM-as-a-Judge ratings, hover text includes the model’s justification.

Clicking \textbf{Examine Viz Grammar Differences} (Figure \ref{fig:multiformat-outputs} (Right)) opens a JSON spec diff viewer, enabling side-by-side comparison of visualization grammars to trace structural discrepancies like missing filters or mismatched encodings.

\Description{This composite figure shows both high-level summaries and detailed utterance-level results. At the top left is the “Overview Recommendation and Metrics” panel, which recommends the best-performing model–prompt combination (GPT-5 Prompt 1 in this case) and shows a matrix of visualization and natural language metric scores. To the right, a “Metrics by Labels” panel slices results by chart type, ambiguity, context handling, and turn numbers, with grouped bar charts showing performance across categories. At the bottom, utterance-level results are shown: the left panel includes the expected response (a histogram of Profit), and the middle panel shows the GPT-5 Nano Prompt 2 response (a multi-colored histogram of Profit by State). To the right are detailed metric scores, with color-coded blocks indicating low, medium, or high performance. This figure connects high-level aggregated metrics to concrete examples, making differences interpretable and actionable.}

This combination of overview, detail, and contextual explanations transforms the metric table into an interactive evaluation workspace, enabling practitioners to interpret scores, and understand how and why they were assigned.

\subsubsection{Overview Panel} \textbf{[D5-6]} 
The Overview panel provides an entry point into evaluation by surfacing progress, recommendations, and aggregate insights. A real-time progress bar tracks completion across models, prompts, and test cases. To mitigate the risk that headline scores anchor practitioners on partial or unstable results, the overview panel is made visible once all utterance-level evaluations for a given run have completed. Each recommendation card links back to the underlying metric table and per-utterance views, encouraging users to treat the overview as a starting hypothesis rather than a definitive judgment. Once evaluation is complete, the system highlights a data-driven recommendation for the best-performing model–prompt pair, based on aggregated metrics. 

To support interpretability, the panel includes \textbf{Overview Metric Cards} (Figures \ref{fig:setupEval}, \ref{fig:multiformat-outputs}) summarizing performance across three key dimensions: visualization, natural language, and traditional NLG metrics. Each card supports drill-down inspection, enabling practitioners to trace how high-level scores emerged from individual test case dimensions, mirroring the overview + detail pattern in the hierarchical metrics table. 

The panel also features a \textbf{Metrics-by-Label} view (Figure~\ref{fig:multilevel-metrics}), which breaks down results by test case annotations such as chart type (e.g., bar, line, scatter), ambiguity class (semantic, syntactic, pragmatic), and contextual intent (e.g., slot-filling, reference resolution, filter carryover). This layered, faceted view helps practitioners move from global trends to specific breakdowns, clarifying \textit{why} a model–prompt pairing was recommended, and where it succeeds or fails.

\subsubsection{Implementation}
The \lexara~toolkit is implemented as a distributed web application with a React frontend \cite{react} using TypeScript \cite{typescript}, Ant Design components \cite{antdesign}, and a Flask backend \cite{flask}. The system follows a microservices architecture with asynchronous job processing and real-time streaming. A Redis Queue library handles background tasks~\cite{rq}, while connection pooling and semaphore-based concurrency control manage API rate limits and prevent resource contention. 

\subsubsection{Toolkit Deployment}
The toolkit is deployed at \url{https://lexara-6b38293fcdac.herokuapp.com/} and has been iteratively refined based on feedback from engineers, designers, PMs, and researchers across multiple CVA teams at a large technology company. It is also available as an open-source project on GitHub \url{https://anonymous.4open.science/r/Lexara-CVA-Eval-280B/README.md} to support broader adoption and experimentation within the CVA research and practitioner communities.

\section{Field Deployment Diary Study: Method \& Findings}
\label{sec:method}
To explore how practitioners use \lexara~in real-world settings, we deployed \lexara within a large technology company that develops a range of CVA products, and recruited a subset of the CVA tool developers from our earlier formative study cohort to participate in a two-week structured diary study [\S\ref{sec:formative}.2].

\begin{table*}[!ht]
\begin{tabular}{@{}
>{\columncolor[HTML]{FFFFFF}}l 
>{\columncolor[HTML]{FFFFFF}}r 
>{\columncolor[HTML]{FFFFFF}}r 
>{\columncolor[HTML]{FFFFFF}}r 
>{\columncolor[HTML]{FFFFFF}}r @{}}
\toprule
\textbf{PID} &
  \multicolumn{1}{l}{\cellcolor[HTML]{FFFFFF}\textbf{\# of Evaluation Experiments}} &
  \multicolumn{1}{l}{\cellcolor[HTML]{FFFFFF}\textbf{\# of Test Cases Executed}} &
  \multicolumn{1}{l}{\cellcolor[HTML]{FFFFFF}\textbf{\# of Models}} &
  \multicolumn{1}{l}{\cellcolor[HTML]{FFFFFF}\textbf{\# of System Prompts}} \\ \midrule
\textbf{P1} & 5  & 62  & 4 & 3 \\
\textbf{P2} & 10 & 103 & 10 & 2 \\
\textbf{P3} & 3  & 16  & 5 & 1 \\
\textbf{P4} & 8  & 45  & 5 & 1 \\
\textbf{P5} & 2  & 32  & 4 & 2 \\
\textbf{P6} & 10 & 85  & 10 & 2 \\ \bottomrule
\end{tabular}
\caption{Summary of evaluation experiments conducted by participants ($P1$–$P6$) during the two-week diary study. Each participant ran multiple experiments across varying test cases, models, and prompts. These figures provide a quantitative overview of how \lexara~was appropriated in practice, complementing our qualitative insights.}
\label{tab:participant_summary}
\end{table*}

\subsection{Study Setup}
We recruited six CVA tool developers (two engineers, one designer, and three product managers) with prior experience evaluating LLMs and prompt strategies for visualization tasks. Before the study, participants joined a 30-minute orientation introducing \lexara's core features and potential evaluation use cases (e.g., testing prompts, comparing models, authoring test cases).

During the two-week study, participants completed daily evaluation tasks using their own data and prompts, submitting structured logs detailing datasources, test cases, model and prompt selections, rationale, outputs, observations, and confidence levels. A 60-minute debrief interview followed to review experiments and workflows.

We collected orientation and debrief transcripts, daily diary logs, \lexara~exports, and participant-authored test cases. All materials were pseudonymized. We conducted thematic analysis using a hybrid inductive–deductive coding approach \cite{charmaz2014constructing} to identify patterns in real-world evaluation practices and tool gaps.

We now report findings from the two-week diary study with six participants ($P1$–$P6$). Our goal was to understand how practitioners used \lexara~in their daily evaluation workflows, what value they derived from its test cases, metrics, interactive features, and remaining gaps. Participants conducted $38$ evaluation experiments across $57$ uniquely authored test cases (rest from existing test case suite), comparing $10$ LLMs and $6$ system prompts (see Table \ref{tab:participant_summary}). 

\subsection{\lexara's Test Cases Captured Real CVA Use} 
Participants valued the realism and variety in the curated test cases, especially the inclusion of multiple chart types and multi-turn follow-ups that mirrored real CVA workflows. As $P6$ noted, ``\textit{one prompt and then the next, then remove nulls, felt like how someone might actually interact.}'' The presence of expected reference outputs also helped participants calibrate model responses; $P3$ reflected, ``\textit{It was helpful to have the reference of what the expected output would have been.}''

However, the current YAML-based authoring workflow posed challenges, particularly for participants with non-engineering roles. PMs and designers found it difficult to contribute, with $P1$ remarking, ``\textit{The YAML barrier makes it harder for PMs to contribute new cases.}'' Participants suggested more accessible authoring tools, such as a point-and-click interface to define utterances, labels, and expected outputs without needing to write structured files.

\subsection{\lexara's Metrics Were Nuanced and Interpretable} 
Participants appreciated that \lexara's metrics were not black-box scores but came with drilldowns and on-hover explanations. These features clarified why a score was given, making the results more actionable. As $P2$ noted, ``\textit{Hovering over visual encodings told me it added this extra channel color, which was the fundamental difference.}'' Even those who did not use the hovers extensively still emphasized the value of the metric suite. $P5$ remarked, ``\textit{What I would like to keep [are] the metrics for sure \dots{} this was the part that is missing from other evaluation tools.}''

Participants also valued the hierarchical, collapsible structure of the metrics. $P3$ appreciated being able to expand only the relevant sections, while $P2$ filtered by specific dimensions of interest, such as axis or sort accuracy. The interpretability helped participants connect high-level scores with concrete differences in outputs.

Importantly, participants recognized the value of graded correctness and support for multiple plausible outputs. $P4$ explained, ``\textit{Accuracy isn't just yes or no. Sometimes it's close enough to be useful; other times a valid-looking chart is misleading.}'' Some participants wanted more customization, such as evolving the metrics to better reflect visualization best practices ($P3$), readability or tone ($P4$), or performance measures like latency and cost ($P2$).

\subsection{\lexara Supported Running a Variety of Experiments at Scale}
Participants conducted a variety of evaluation experiments by holding some variables constant while probing others. Common comparisons included large vs. compact models, cross-family competitors, and different prompting strategies, such as persona-prompting (e.g., ``\textit{as a data visualization expert}'' $P3$), few-shot examples (e.g., ``\textit{learn color coding based on these more engaging visualizations}'' $P6$), or prompts in different languages to test tone and formality ($P4$). Some explored edge cases, like how multi-turn interactions handled filters and sorts.

Workflows generally followed an overview-to-detail pattern aligned with \lexara's design: selecting test cases, running model-prompt combinations, reviewing summary metrics and system recommendations, inspecting rendered outputs, and drilling into JSON diffs. $P2$ valued the concise summary: ``\textit{I appreciate the top-level recommendation}.'' P4 praised the structured comparison: ``\textit{Great for like-for-like comparisons.}'' $P6$ added, ``\textit{It's cool to see the model outputs side by side and compare how they generated the viz.}'' Some participants ran parallel experiments in separate tabs to test multiple hypotheses simultaneously.

\subsection{\lexara Facilitated Granular, Multi-Format Evaluation}
\lexara~enabled participants to inspect model differences across multiple granularities, formats, turns, and runs, aligning with design goals ($D4$–$D6$). The side-by-side view of expected and actual outputs emerged as the most intuitive entry point. $P3$ appreciated that it surfaced divergences clearly: ``\textit{I liked the side-by-side, and having the notional spec JSON to see exactly where differences came from.}'' $P6$ emphasized it made follow-up failures obvious: ``\textit{Seeing it next to the reference chart made that obvious.}'' Compared to spreadsheets or tab-switching, participants found this interface reduced cognitive load.

The JSON diff viewer added a deeper diagnostic lens, helping explain score mismatches when charts looked similar. $P3$ noted: ``\textit{Two vizzes looked the same, but the score wasn't 100. JSON showed a tooltip difference.}'' They used it to uncover hidden mismatches in encodings: ``\textit{Empty graphs still got high scores, but the axis binding was off.}'' In one sequence, $P1$ initially tested GPT-4o-mini but noticed misaligned encodings when inspecting results. \lexara's JSON diff viewer highlighted the mismatch, prompting them to switch to Claude Opus 4, saying ``\textit{The Viz Grammar Diff is pretty handy!\dots{} More confident than before! The difference in scores does seem to correlate better with the observed differences in the viz response.}'' 

Participants also used the overview metrics and recommendation cards as starting points. While some appreciated the concise summaries ($P2$: ``\textit{I appreciate the top-level recommendation}''), others cross-checked them with their own assessments. $P6$ challenged a suggestion: ``\textit{It recommended the mini, but my tally favored another.}'' $P1$ noted subtle visual flaws not reflected in the metrics. P1 said, \textit{``I feel quite confident. I saw quantitatively a stark difference in the performance of the models and also by clicking through the outputs could tell qualitatively that Claude Sonnet was matching the expected outputs more often.''} While initial use presented a learning curve ($P4$: ``\textit{The breakdown was almost too much at first}''), participants ultimately integrated the interface into sensemaking workflows, validating model and system prompt outputs, interpreting discrepancies, and reasoning across CVA’s multi-format outputs.

\section{Validating \lexara's Metrics}
\label{sec:metricevaluation}
To assess whether \lexara's metrics align with expert judgment, we conducted a quantitative validation study comparing metric outputs against human ratings of CVA responses.

\subsection{Method}
We sampled $N = 120$ CVA responses from the formative and diary study experiments (\S3, \S6), stratified with coverage across: different metrics and score ranges (lower, medium, higher thirds of scores); different ambiguity labels (e.g., syntactic, semantic, pragmatic); different task types (e.g., descriptive vs. comparative vs. trend analysis). Each sampled response included: the datasource schema, the user utterance (and conversational context where applicable), and the model-produced visualization, JSON specification, and natural language response.

Two raters (R1, R2), diary study participants familiar with \lexara scored each CVA response on all the metrics defined in \S\ref{sec:metrics}, using the native scale associated with that metric. Raters completed a training phase of 10 pilot items, where they could clarify how to interpret the rubric with the authors, but not discuss individual items or specific scores. These pilot items were not considered in the final analysis.

For each metric, we computed (1) inter-rater reliability between the two raters using linear-weighted Cohen’s $\kappa$, quantifying how consistently raters applied the rubric to the same set of responses and (2) metric–human alignment between the mean human score per response (simple average of the two raters) and metric, and then calculated Spearman’s rank correlation $\rho$.

\subsection{Results}
\subsubsection{Inter-Rater Reliability} Human raters showed moderate to high agreement on most metrics (see Appendix Figure \ref{fig:irr}). Across visualization metrics, linear-weighted $\kappa$ ranged from $0.45$ to $0.78$ (median $\kappa$ = 0.65), highest for Data Fidelity, Field, Chart Type, and Axis, Filter, Sort Accuracy, and lowest for Interactivity, reflecting the greater subjectivity of interaction design. For natural language response metrics, $\kappa$ ranged from $0.46$ to $0.80$ (median $\kappa$ = 0.63), with Factual Grounding and Coherence exhibiting higher agreement than Insightfulness and Follow-up Relevance. These results suggest that human raters can apply \lexara's metrics reliably. Some fluctuation is expected in metrics that capture subjective or experiential qualities like interactivity judgments as they depend on evaluators’ expectations about analytic workflows, prior tool experience, and task context. Rather than treating these metrics as decisive indicators of overall system quality, we recommend interpreting them as diagnostic signals. These metrics are intended to complement, not override, more objective correctness measures, guiding targeted debugging and design iteration rather than serving as pass or fail criteria.

\subsubsection{Metric–Human Correlation} \lexara's  metrics aligned well with human judgments: Data Fidelity, Field Similarity, and Chart Type Similarity showed strong rank correlations with human ratings (Spearman’s $\rho$ in the range $0.68–0.79$, see Appendix Figure \ref{fig:metric-human}). In natural language response metrics: Factual Grounding exhibited the strongest alignment ($\rho$ = 0.82). The remainder of the natural language response metrics correlated at  ($\rho$ = 0.57–0.71, see Appendix Figure \ref{fig:metric-human}), lower than Factual Grounding but comparable to human–human agreement. 

While Lexara incorporates multiple safeguards to reduce LLM-as-a-Judge biases, disagreements between automated judges and human evaluators still occur, particularly on subjective dimensions. For example, in one evaluation instance, a model-generated bar chart included correct data mappings and filters but omitted interactive tooltips. Human raters penalized this omission due to its impact on exploratory analysis, whereas the automated judge assigned a relatively high interactivity score based on the presence of a rendered chart and valid specification structure. Lexara surfaces such disagreements explicitly in the interface by exposing per-metric scores, judge rationales, and underlying visualization specifications alongside rendered outputs. This allows practitioners to inspect where and why judgments diverge, override automated scores when appropriate, and treat automated evaluations as assistive rather than authoritative. By supporting this human-in-the-loop workflow, Lexara enables users to balance scalability with contextual judgment, reinforcing trust in the evaluation process rather than obscuring uncertainty.

\subsubsection{Model Alignment with Human Preferences} For each of the ten LLMs evaluated in the diary study (\S6.1), we computed the mean score by averaging all visualization metrics and all natural language response metrics across all test cases and system prompts that participants executed with that model. Independently, at the end of the diary study we had asked participants for the rankings of each model they had interacted with: (i) an overall 1–5 quality rating for CVA tasks and (ii) a rough rank ordering of models from best overall for CVA to worst (allowing ties). We converted these into per-model human preference scores by (a) normalizing each participant’s ranks to [0, 1], (b) averaging across participants for each model (ignoring models a participant had not used), and (c) using these averages as the human judgment baseline. We then computed Spearman rank correlations between each model's human preference score and its mean visualization and natural language response score, quantifying how well the model performance aligns with practitioners' preferences. Models that participants perceived as stronger for CVA tasks generally obtained higher mean scores across both visualization and natural language response metrics (see Appendix Figures \ref{fig:model-viz}, \ref{fig:model-nl}). The rank correlation between human preferences and \lexara's overall visualization score was $\rho$ = 0.79 (p < 0.01), and $\rho$ = 0.74 (p < 0.05) for the natural language response score. These results do not constitute a full comparative benchmark, but they provide a sanity check that \lexara's metrics track practitioners’ qualitative impressions at the coarse model level, complementing the per-metric validation against expert ratings.

\section{Limitations and Future Work}
\label{sec:discussion}
\lexara~contributes to a growing body of research on evaluating LLMs, with a particular focus on the unique demands of CVA. Prior work has offered important building blocks: large-scale text benchmarks for reasoning and language quality~\cite{Liang2022HolisticEO,zhu2024promptbench}; visualization-specific test suites~\cite{luo2021nvbench,viseval}; and interactive LLM evaluation toolkits~\cite{Arawjo2024ChainForge,Kim2024EvalLM,Kahng2024LLMComparatorEA}. However, these efforts typically focus on single-turn, text-only outputs or require significant programming effort to evaluate, making them less suitable for evaluating multi-format, multi-turn, and ambiguity-rich CVA workflows. \lexara~addresses these limitations by integrating interpretable metrics, grounded real-world test cases, and an accessible low-code interface tailored for CVA evaluation. 

\subsection{Broadening Scope for Sustained Use of LLM-based CVA Evaluation Toolkits}
While our diary study demonstrates \lexara's usefulness for systematic LLM evaluation, several limitations suggest future directions. The test suite coverage, though designed for multi-turn, multi-format CVA conversations, remains bounded by datasources, domains, and intents from our formative studies and existing benchmarks. The toolkit currently assumes access to expected CVA responses for each test case, reflecting its diagnostic benchmarking role. Extending the interface for ad hoc exploratory use remains an open design challenge. \lexara~currently supports common chart types (bar, line, scatter, histogram, box plot, multivariate line, pie charts). Building on Vega-Lite's expressive grammar, the toolkit is extensible to broader visualizations (maps, Sankey diagrams, heatmaps) by authoring new test cases using declarative Vega-Lite specifications. Additional contributions may uncover new test cases as conversational intents evolve from analytic questions to dashboard authoring or data stories.

Broader adoption could reveal how sustained use impacts trust, model selection, and deployment practices. Future iterations should expand the test suite across more domains, user types, and data modalities. As users upload custom test cases, features should support ethically grounded, opt-in contribution mechanisms, raising questions around consent, credit, and data quality.

The current YAML/JSON authoring workflow poses challenges for non-technical stakeholders despite offering transparency and control. The diary study revealed desire for easier pathways (CSV templates, point-and-click builders), but simplification risks sacrificing precision and reproducibility critical for benchmarking. A promising direction is collaborative authoring: engineers specify formal test logic while designers and analysts contribute natural language utterances and qualitative labels, aligning with HCI research on participatory evaluation and mixed-expertise workflows. This work does not address operational concerns like cost, latency, prompt/model drift-critical for large-scale deployment. Incorporating these aspects could enable more holistic, real-world CVA tool evaluations. We have open-sourced the project and hope the community continues to develop this.

A limitation of our validation of LLM-as-a-Judge metrics is that the human raters were \lexara-experienced and trained by the authors. While this familiarity may bias judgments toward the toolkit’s rubrics and increase alignment with automated metrics, we intentionally adopted this setup for an initial quantitative validation to reduce labeling noise when applying nuanced, graded CVA criteria; this is reflected in high inter-rater reliability (Cohen’s $\kappa$ = 0.81). Establishing this calibrated baseline allows us to characterize metric behavior before introducing additional sources of variance. Future work should evaluate generalizability by involving independent domain experts and blinded crowd raters, comparing inter-group agreement, and using randomized and double-blind rating protocols to detect systematic bias and assess transfer beyond this expert-curated setting.

\subsection{Designing CVA Metrics for More Nuanced Evaluation Strategies}
\lexara~extends beyond existing visualization benchmarking efforts~\cite{podo2024vievallmconceptualstack, dibia2023lida, VISShepherd2025}, by introducing user-centered, graded metrics that move beyond binary checks of validity, legality, and readability. These include finer-grained measures of visualization specification fidelity (e.g., field, axis, sort accuracy) and natural language response quality (e.g., insightfulness, grounding), designed to support multi-format, multi-turn CVA tasks. However, several limitations remain. 

To focus on evaluating the baseline capabilities and limitations of LLMs, our current metrics evaluate model outputs derived from text prompts and JSON specifications; they do not yet assess models' native multimodal perception of rendered charts or their performance when using external tools.

Evaluation metrics often encode subjective judgments: thresholds for specification similarity and heuristics for matching may reflect implicit normative biases\cite{ shi2024judging,wataoka2024self}. Over-optimization toward these metrics could obscure genuine analytic quality. We view the scores as providing a baseline check on whether CVA outputs are eligible to support analytic reasoning (e.g., correct data, appropriate encodings, factually grounded explanations), rather than as direct proxies for the quality of the human sensemaking process itself. While \lexara~visualizes full conversational sequences through the metrics-by-label and drill-down views, we do not yet provide explicit trajectory measures such as the number of turns required to reach an acceptable chart or the frequency of successful self-correction.

Ecological validity also introduces variability. Real-world utterances yield multiple plausible answers, complicating reproducibility and inter-rater consistency. While \lexara's graded metrics offer partial interpretability, evaluating under ambiguity remains a broader methodological challenge in HCI and NLP evaluation~\cite{battle:2018, belz-etal-2023-non, clark2021all}.

\subsection{Supporting Actionable Sensemaking in CVA Benchmarking Workflows}
Our work combines interactive visualization rendering, JSON spec diffs, hierarchical metric breakdowns, and progress overviews to support more nuanced diagnosis of model and prompt configurations. However, the downside of these interface enhancements is the learning curve; rich outputs can overwhelm new users, and auto-generated recommendations occasionally diverge from practitioners' qualitative judgments, prompting additional manual review. YAML-based authoring also persists as a bottleneck, and integration with enterprise tools for collaboration, orchestration, or data authoring remains limited.

More fundamentally, \lexara~functions as a diagnostic CVA benchmarking toolkit; it reveals \textit{where} and \textit{why} models fall short, but does not yet close the loop to support more actionable sensemaking. Future extensions could support semi-automated prompt repair~\cite{zhou2023lima, peng:2023} or training data augmentation~\cite{Zhou2022LeasttoMostPE} based on failure patterns, transforming evaluation from a retrospective analysis into a forward-looking, feedback-driven improvement loop. This would align evaluation more closely with iterative development workflows~\cite{zhu2024promptbench,Liang2022HolisticEO}.

\section{Conclusion}
\label{sec:conclusion}
As LLMs increasingly mediate analytical reasoning and visual exploration, rigorous and user-centered evaluation becomes critical. Through formative studies with practitioners, we identified key challenges in evaluating LLMs for CVA, including test cases misaligned with real-world use cases, a lack of interpretable, graded metrics, and ad hoc fragmented evaluation workflows. We operationalize these insights into \lexara, a user-centered CVA evaluation toolkit including test cases grounded in real-world CVA use cases, interpretable metrics that account for multiple or partially correct responses, and supporting low-code benchmarking balancing human and automatic evaluation methods. By enabling scalable, nuanced, and CVA-specific evaluation, our work contributes both conceptual and practical advances toward more transparent, trustworthy, and user-centered assessment of LLM behavior in CVA systems. The toolkit is publicly available at \url{https://lexara-6b38293fcdac.herokuapp.com/} with open-source code at \url{https://anonymous.4open.science/r/Lexara-CVA-Eval-280B/README.md}, to support broader adoption and extension by the HCI and visual analytics communities.


\bibliographystyle{ACM-Reference-Format}
\bibliography{main}
\newpage

\label{sec:appendix}
\section*{Appendix}
\textbf{Formative Study Apparatus}
\begin{figure}[!ht]
\centering
\includegraphics[width=\columnwidth]{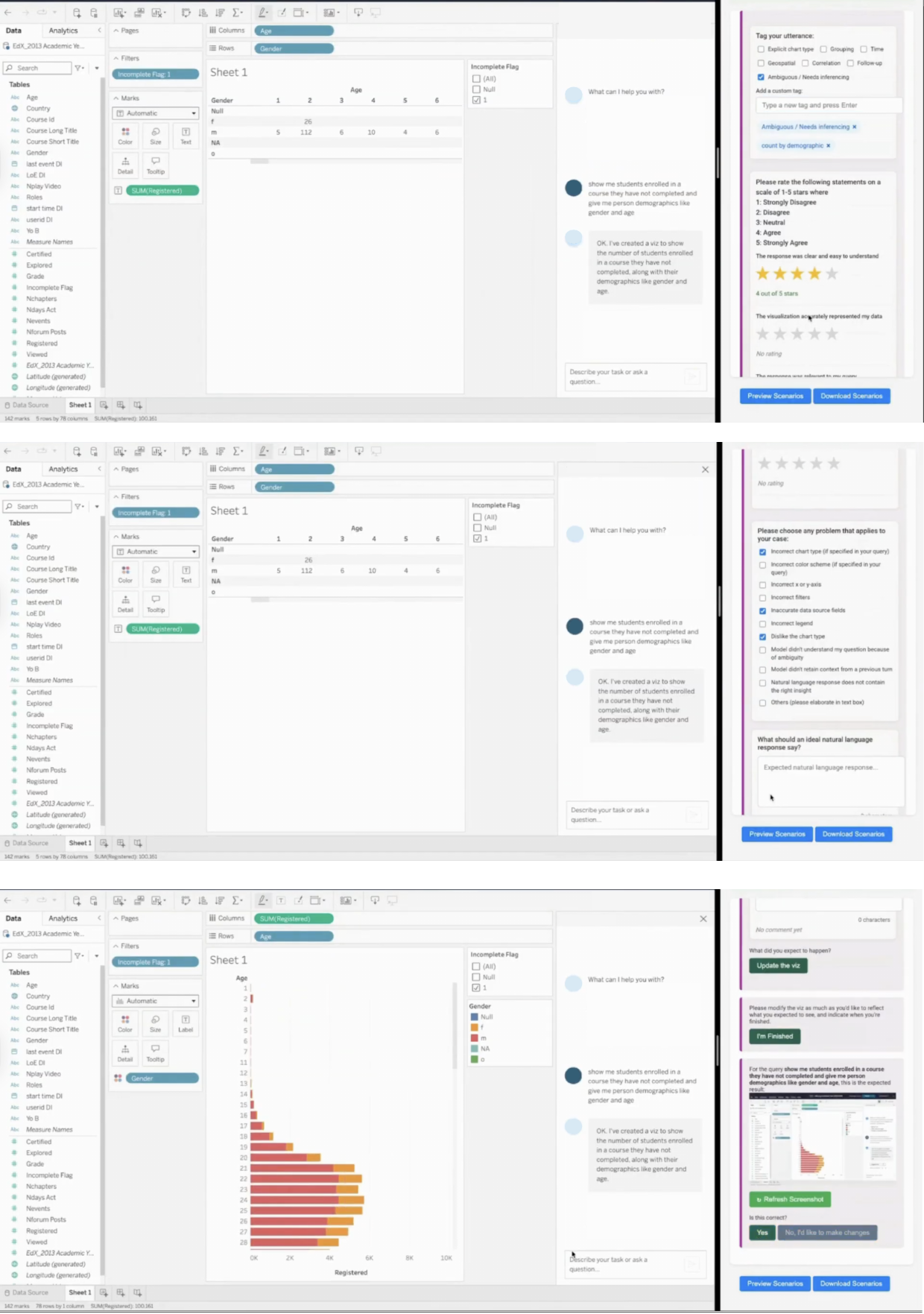}
\caption{A browser plugin recorded participants' interactions with a popular CVA tool, capturing their utterances, model responses, in-the-loop evaluations via Likert-type scales, and corrected expected outputs.}
\label{fig:tc-extension}
\Description{Screenshot series of a browser plugin recording participants' interactions with a conversational visual analytics tool. On the left, the tool's main interface shows data fields, filters, and a chart canvas in Tableau Agent. On the right, the plugin panel logs the participant’s utterances, model responses, Likert-type scales, and fields for corrections. The first screenshot shows a simple text table with minimal charting; the second shows a partially completed bar chart; the third shows a fully rendered stacked bar chart with colored bars by region. The side panel consistently shows user utterances, system responses, and rating sliders.}
\end{figure}

\begin{table*}[ht]
\centering
\footnotesize
\begin{tabularx}{\textwidth}{@{}>{\raggedright\arraybackslash}p{4.5cm} >{\raggedright\arraybackslash}p{2.5cm} >{\raggedright\arraybackslash}X >{\raggedright\arraybackslash}p{3cm}@{}}
\toprule
\textbf{Datasource} & \textbf{Domain} & \textbf{Details} & \textbf{Source} \\
\midrule
\href{https://www.kaggle.com/datasets/vivek468/superstore-dataset-final}{Superstore} & Business \& Finance & Contains information about products, sales, and profits that can be used to identify key areas of improvement within this fictitious company. & Tableau \\
\href{https://www.kaggle.com/datasets/prashant808/the-2014-inc-5000-list}{The 2014 Inc. 5000} & Business \& Finance & Inc. Magazine's annual list of the 5000 fastest growing private companies in the U.S., compiled by measuring each company's percentage revenue growth over a four-year period. & Inc. Magazine \\
\href{https://www.kaggle.com/datasets/prashant808/the-top-paying-sports-teams-and-top-paid-athletes}{Global Sport Finances} & Business \& Finance & The top-paying pro sports teams and the top paid athletes. & ESPN \\
\href{https://www.kaggle.com/datasets/sumithbhongale/american-university-data-ipeds-dataset}{American University Data (IPEDS)} & Education & Primary source for data on colleges, universities, and technical/vocational postsecondary institutions in the U.S. & National Center for Education Statistics \\
\href{https://dataverse.harvard.edu/dataset.xhtml?persistentId=doi:10.7910/DVN/26147}{edX/HarvardX (AY 2012--2013)} & Education & De-identified data from the first year (Fall 2012, Spring 2013, Summer 2013) of MITx and HarvardX courses on the edX platform. & Harvard Dataverse \\
\href{https://www.kaggle.com/datasets/kumarajarshi/life-expectancy-who}{Life Expectancy WHO} & Healthcare & Historical and current life expectancy by country, often paired with other health indicators. & World Health Organization \\
\href{https://data.who.int/dashboards/covid19/data?n=o}{Global Vaccination Coverage for COVID-19} & Healthcare & Tracks immunization coverage for COVID-19 vaccines across different countries over multiple years. & World Health Organization
\end{tabularx}
\caption{Overview of datasources across business \& finance, education, and healthcare domains, each linked to its original source.}
\label{tab:formative-datasets}
\end{table*}

\begin{figure*}[!ht]
\includegraphics[width=\textwidth]{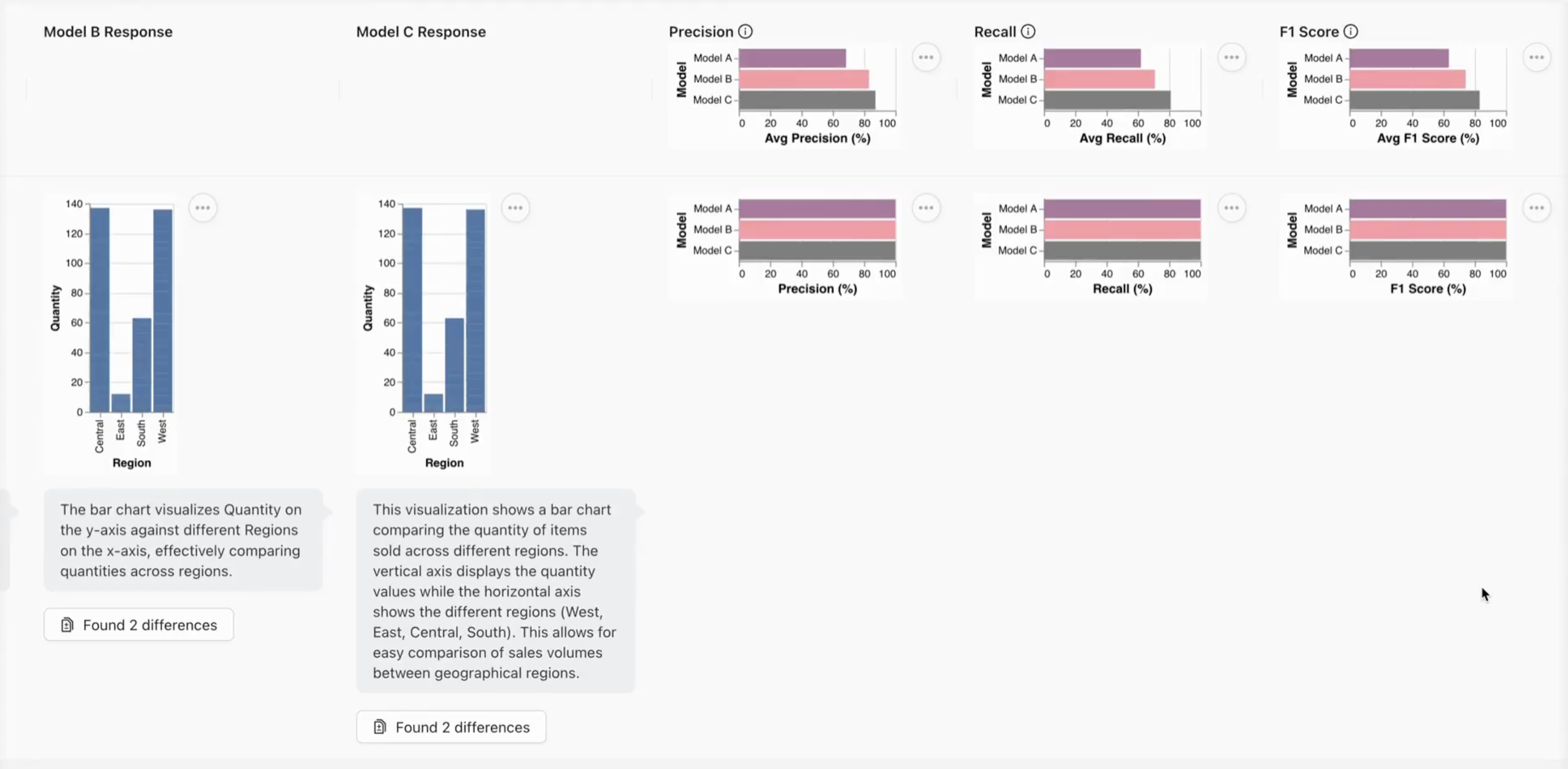}
\caption{In the formative study, participants reviewed three anonymized model outputs presented in a static table, alongside traditional NLG metrics including F1, Precision, Recall for each response.} 
\label{fig:model-diffs}
\end{figure*} 
\Description{A static evaluation table comparing three anonymized model outputs side-by-side for the same utterance. Each column contains bar charts of ``Quantity by Region'' along with textual descriptions. Below each output are difference markers showing where outputs diverged from the reference. Beneath the charts, traditional NLG metrics (Precision, Recall, F1 score) are displayed as small bar charts for each model response. The figure demonstrates how participants reviewed outputs with both visualization and text metrics in parallel.}

\begin{figure*}[t]
    \centering
    \begin{subfigure}{0.48\linewidth}
        \centering
        \includegraphics[width=\linewidth]{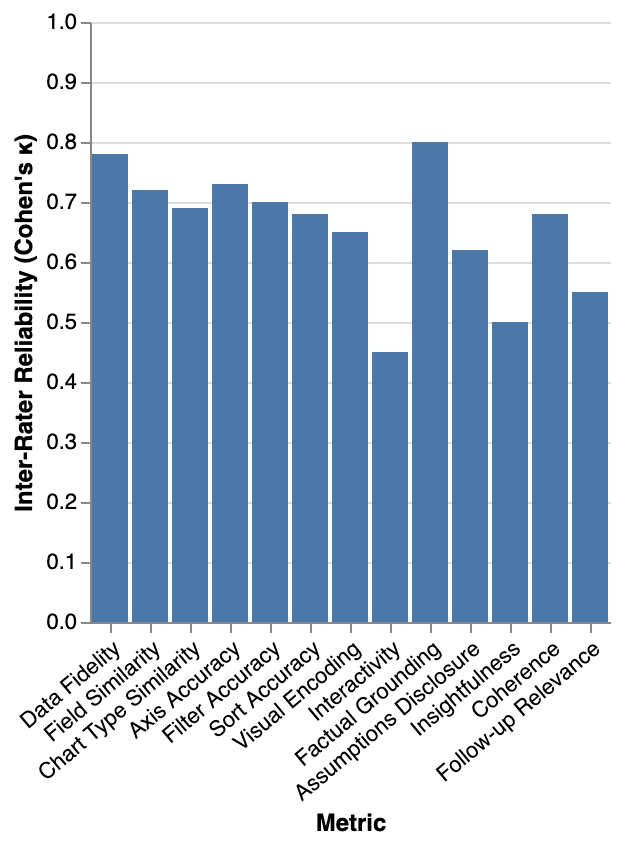}
        \caption{Inter-rater reliability (Cohen's $\kappa$) for all metrics.}
        \label{fig:irr}
    \end{subfigure}\hfill
    \begin{subfigure}{0.48\linewidth}
        \centering
        \includegraphics[width=\linewidth]{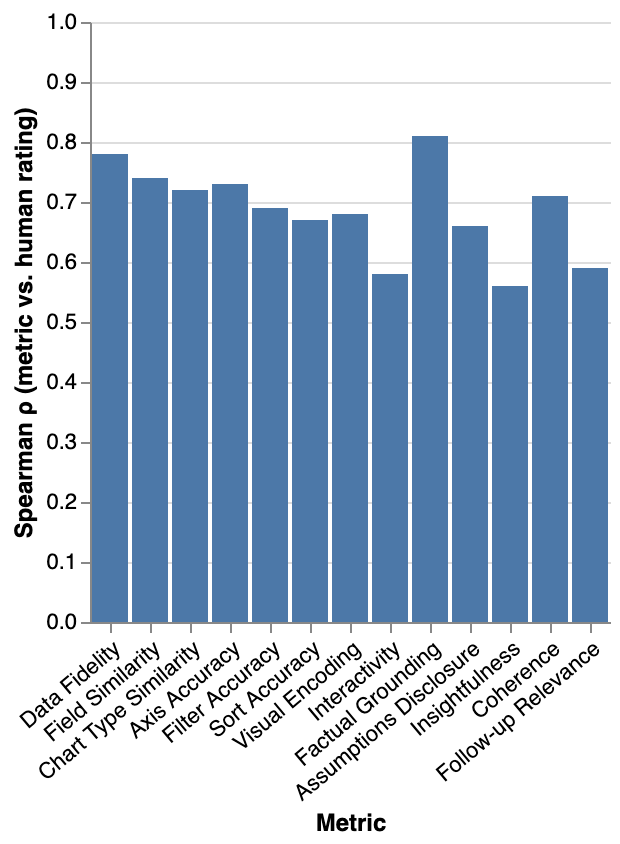}
        \caption{Metric–human alignment (Spearman $\rho$) for all metrics.}
        \label{fig:metric-human}
    \end{subfigure}

    \vspace{0.8\baselineskip}

    \begin{subfigure}{0.48\linewidth}
        \centering
        \includegraphics[width=\linewidth]{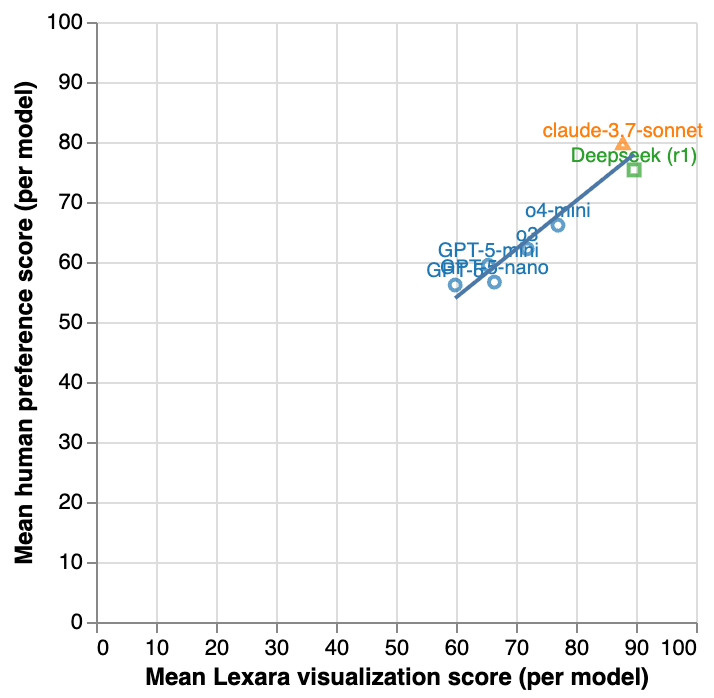}
        \caption{Model-level alignment for visualization scores.}
        \label{fig:model-viz}
    \end{subfigure}\hfill
    \begin{subfigure}{0.48\linewidth}
        \centering
        \includegraphics[width=\linewidth]{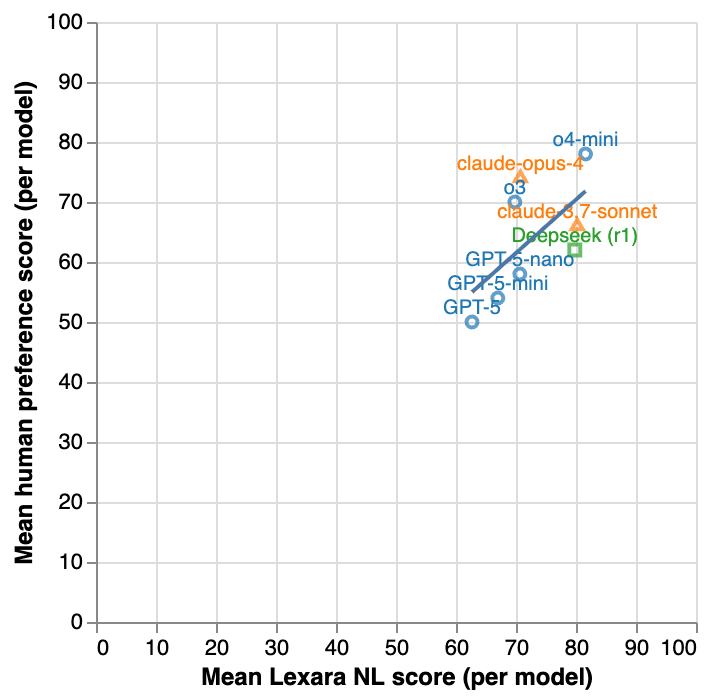}
        \caption{Model-level alignment for natural language/conversation scores.}
        \label{fig:model-nl}
    \end{subfigure}

    \caption{Lexara metrics are both reliable and aligned with human judgments: (a–b) most metrics show $\kappa$ and $\rho$ above $0.6$; (c–d) models with higher Lexara visualization and natural language scores are also preferred by humans.}
    \label{fig:lexara-validation}
\Description{Figure comprises four panels validating Lexara's evaluation metrics against human judgment.
Panel (a) shows a bar chart of inter-rater reliability measured by Cohen's kappa across 13 metrics. Most metrics show kappa values between 0.45 and 0.80, indicating moderate to high agreement. The highest reliability (approximately 0.78-0.80) is observed for Data Fidelity, Field Similarity, and Chart Type Similarity. Factual Grounding also shows high reliability around 0.80. Interactivity shows the lowest reliability at approximately 0.45. The median kappa across all metrics is 0.65.
Panel (b) displays a bar chart of metric-human alignment using Spearman's rho correlation. Values range from approximately 0.57 to 0.82. Factual Grounding achieves the strongest alignment at 0.82. Data Fidelity, Field Similarity, and Chart Type Similarity show correlations between 0.68 and 0.79. Natural language metrics including Insightfulness, Coherence, and Follow-up Relevance show moderate correlations between 0.57 and 0.71.
Panel (c) presents a scatter plot comparing mean Lexara visualization scores (x-axis, 0-100) against mean human preference scores (y-axis, 0-100) for ten LLM models. Models are labeled and color-coded, including claude-3.7-sonnet and Deepseek (r1) at the upper right (high on both metrics), o4-mini, GPT-5-mini, and GPT-5-nano in the middle range, and GPT-5 at the lower left. The positive correlation (Spearman ρ = 0.79, p < 0.01) indicates models scoring higher on Lexara visualization metrics are also preferred by human evaluators.
Panel (d) shows a similar scatter plot for natural language and conversation scores. The x-axis represents mean Lexara natural language scores (0-100) and y-axis shows human preference (0-100). o4-mini and claude-opus-4 appear in the upper right quadrant, claude-3.7-sonnet and Deepseek (r1) in the middle-upper range, and GPT-5, GPT-5-mini, and GPT-5-nano cluster in the lower left. The correlation is ρ = 0.74 (p < 0.05), demonstrating that Lexara's natural language quality metrics align with practitioner preferences.
Overall, the figure demonstrates that Lexara's automated metrics are both internally reliable (consistent between human raters) and externally valid (aligned with human quality judgments), supporting their use for systematic CVA evaluation.}
\end{figure*}


\end{document}